\documentclass[aps,prb,twocolumn,amsmath,amssymb,superscriptaddress,floatfix] {revtex4}
\usepackage{amsmath,amsfonts,amsthm,graphicx}
\usepackage{graphics}
\usepackage{dcolumn}
\usepackage{bm}
\newcommand{\beq}{\begin{equation}}
\newcommand{\eeq}{\end{equation}}
\newcommand{\beqnar}{\begin{eqnarray}}
\newcommand{\eeqnar}{\end{eqnarray}}
\newcommand{\bfig}{\begin{figure}}
\newcommand{\efig}{\end{figure}}

\usepackage{multirow}
\usepackage{graphicx}
\usepackage{diagbox}

\usepackage{epsfig}
\usepackage{epstopdf}
\usepackage{grffile}
\usepackage[english]{babel}
\usepackage{bm}
\usepackage{sidecap}
\usepackage{array}
\usepackage{romannum}
\usepackage{float}

\usepackage{lipsum}
\usepackage{natbib}

\newcommand{\comment}[1]{\textcolor{red}{#1}}
\renewcommand{\comment}[1]{\relax}
\newcommand{\tobedeleted}[1]{\textcolor{green}{#1}}
\renewcommand{\tobedeleted}[1]{\relax}

\begin{document}

\title{Straintronics in graphene nanoribbons}
\author{Maliheh Azadparvar}
\affiliation{School of Physics, Damghan University, P.O. Box 36716-41167, Damghan, Iran}
\author{Hosein Cheraghchi}
\email{cheraghchi@du.ac.ir}
\affiliation{School of Physics, Damghan University, P.O. Box 36716-41167, Damghan, Iran}
\affiliation{Physics Department, Iran University of Science and Technology, P. O. Box, 16844, Narmak, Tehran, Iran}

\date{\today}

\begin{abstract}
The interplay between uniaxial strain and charging effects in zigzag graphene nanoribbons (ZGNR) is investigated by using non-equilibrium Green's function formalism. The I-V characteristic curves and especially negative differential resistance (NDR) induced by some quantum selection rules are affected by the type, strength and also direction of the applied strain. For the oblique strain, the parity conservation fails while it conserves at the longitudinal and transverse strains. Therefore, in the oblique strain, on-off current ratio drastically decreases in compared to the un-strained case. For the tensile strain along the ribbon axis, $ I_{on}/I_{off} $ ratio increases while for the compressive strain, NDR will be gradually disappeared. This property can be useful for nano-electromechanical switch in which by changing the tensile to compressive strain, current switches between its on and off-current state. Furthermore, under influence of the oblique uniaxial strain, the geometry symmetry of charge accumulation as well as bond local currents will be broken across ZGNR giving rise to a transverse current between upper and lower edges of the nanoribbon. This transverse current would be enhanced by the strain strength where gradient of charge density would be maximum at the off-current state.        
\end{abstract}
\maketitle

\section{Introduction}

Mechanical properties of graphene when are combined with its strange conductivity properties makes graphene one of the potential materials for elastic electronic devices \cite{novoselov2012roadmap,kim2009large,shao2015graphene,strain_review} such as touch screens, electronic papers and foldable organic light-emitting diode. Indeed, graphene which is the strongest 2D material ever measured has ability to be sustain reversible tensile strain as large as $25$ percentage\cite{lee2008measurement}. There are inevitable experimental ways for applying strain in garphene\cite{strain_review}: surface corrugations arising from the substrate\cite{teague2009evidence}, lattice mismatch between graphene and substrate\cite{ni2008raman}, strain induced by ripples and wrinkling\cite{zhang2011maximum,sun2009atomic,bronsgeest2015strain}, mechanical edge warping or twisting instability\cite{huang2009quantum,shenoy2008edge}. Strain also can be applied in a controlled and engineered way such as using a piezoelectric substrate for controllably shrinking or elongating graphene plate by applying a bias voltage\cite{ding2010stretchable}. So controlling graphene properties\cite{falko2014} especially engineering its quantum transport properties is highly possible by using strain, which is so called "straintronics"\cite{straintronics1,straintronics2,straintronics3,straintronics4} leading to quantum strain transistors\cite{PRAppl2019,SciRep2017}. It can be shown that uniaxial and also shear strain can induce a gap in graphene while biaxial strain only changes its Fermi velocity\cite{choi2010effects}.
Furthermore, it is demonstrated that strain on graphene induces the pesudo-magnetic field which generates Landau levels. It has also been predicted the strain-induced quantum hall effect in graphene by the arc-bend strain\cite{chang2012quantum}. 

\begin{figure}[h]
	\centering
	\includegraphics[width=0.55\linewidth] {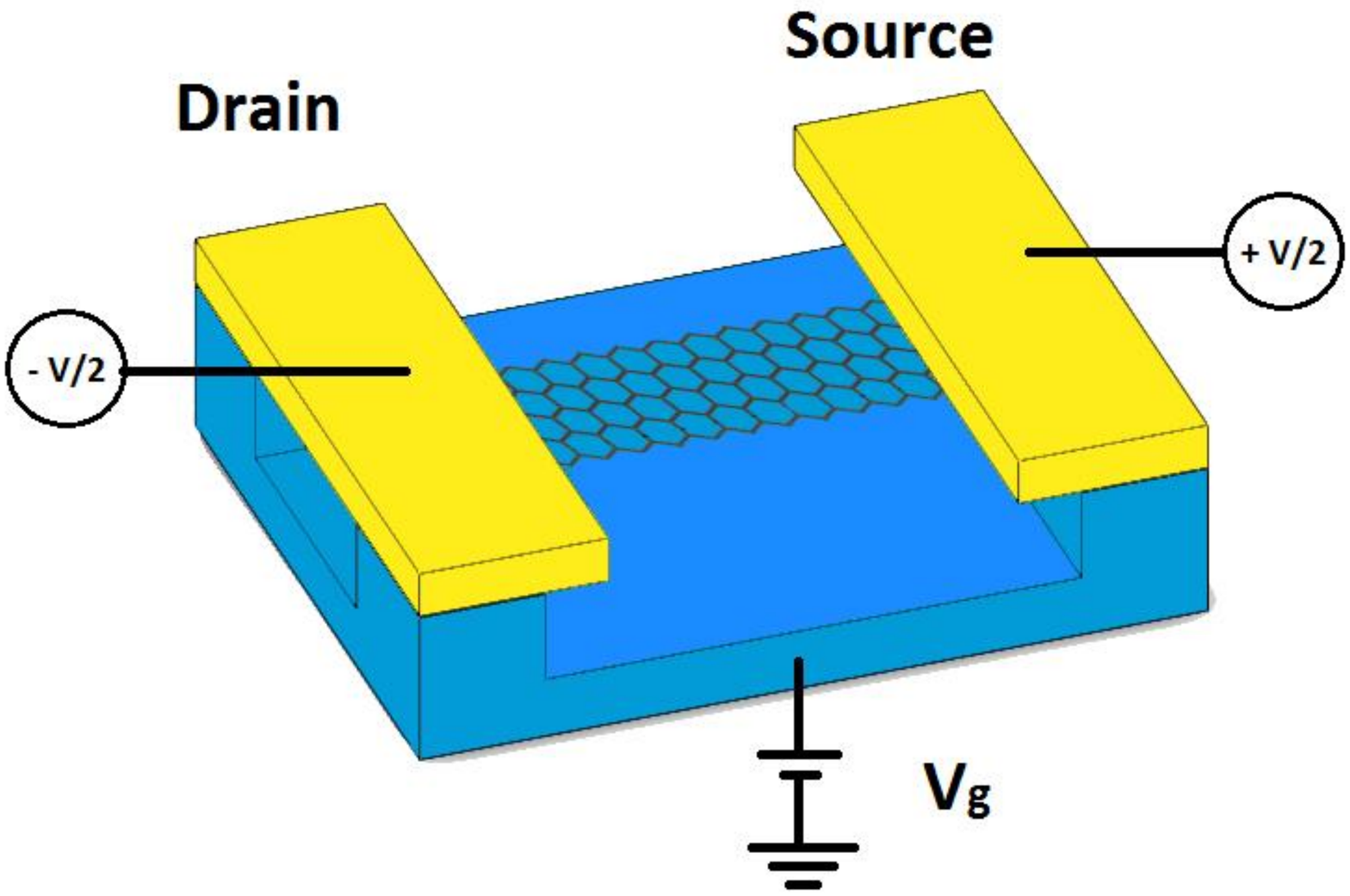}
	\includegraphics[width=0.95\linewidth]{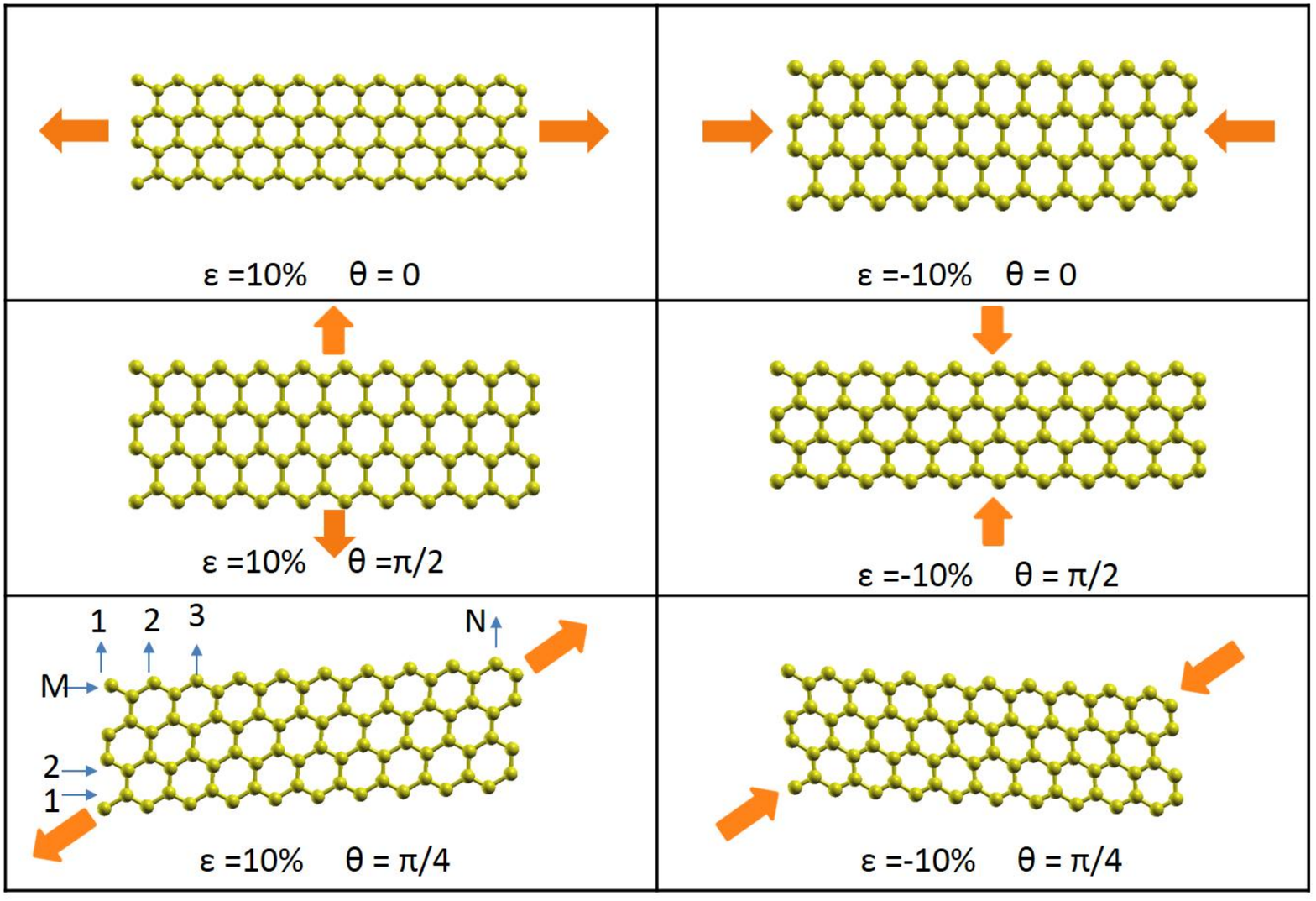}
	\caption{Top panel: Schematic view of a field-effect transistor structure based on zigzag graphene ribbons where the gate voltage is applied on the whole system. Bottom panel: zigzag graphene nanoribbon under uniaxial strain in different directions ($\theta =0, \pi/2$) and ($\theta=\pi/4$).}
	\label{fig:intro}
\end{figure} 

On the other hand, nowadays, experimental methods for synthesis of graphene nanoribbon (GNR) have been improved leading to smooth zigzag or armchair edges. Lithographic methods\cite{han2007energy}, unzipping of carbon nanotubes\cite{jiao2009narrow} and also chemical synthesis\cite{datta2008crystallographic} are some of techniques which have been developed to achieve high quality GNRs with widths lesser than 10 nm. Moreover, by using the bottom-up synthesis, GNR's sub-10 nm field-effect transistors have been demonstrated with a large on/off ratio of current at room temperature\cite{li2008chemically,wang2008room}. Electronic structure of graphene nanoribbons depends strongly on its width and its edge type\cite{Luoie2006}. Zigzag graphene nanoribbons (ZGNRs) show a metallic behavior. However, based on parity selection rule in ZGNR's with even number of zigzag chain in width, one can manipulate a transport gap leading to ZGNR's electronic switch\cite{cheraghchi2010gate,cheraghchi2011nonlinear,Grosso1,Grosso2}. Moreover, prohibition of transport between discontinuous energy bands, governs transport properties of ZGNRs (For more see Appendix. \ref{appendix:a}). These transition rules when are mixed with strain in the presence of e-e interaction, results in interesting and novel phenomena. In carbon nanotubes, there is also a rotational symmetry which governs quantum transport properties as well\cite{rotational}. It is also reported that, at low biases, screening in ZGNR's is effective such that the electrostatic potential is nearly flat inside ZGNRs far from electrodes\cite{cheraghchi2010gate}. However, at high voltages, looking at the charge profile demonstrates that charge is accumulated on the edges which strongly affects the screening. Furthermore, it has been reported the ability of manipulating the voltage drop in graphene nanojunctions by using the gate potential\cite{Brandbyge2015}. Based on {\it ab initio} calculations, it is demonstrated that strong reduction of the effective on-site Coulomb interaction (Hubbard term-$U_H$) at the edges of ZGNRs would be a result of the edge states\cite{hadipour2018screening}. Moreover, the non-local part of interaction would be strongly screened because of these edge states. However, screening in GNR's with armchair edges is non-conventional such that in the intermediate distances there is an anti-screening effect while it is fully screened in short distances\cite{hadipour2018screening}.    

For better understanding of transport properties of ZGNRs under applied strain, it is necessary to know the spatial distribution of charge and electrostatic potential profiles inside the structure. An important question that arises is how charge density and electrostatic potential and also bond current densities are redistributed when an uniaxial strain is applied. In this paper, by using self-consistent (SCF) tight-binding approach, charging effects of ZGNR's is investigated. Negative differential resistance inducing by the quantum selection rules is affected by the tensile or compressive strain. The interplay between transferred charge and current of carriers is presented based on the SCF chagre and potential profiles as well as the bond currents among ZGNR. It is demonstrated that in the oblique strains, there is a relation between imbalanced charge at the edges and also longitudinal current at the on or off-current states. Imbalanced charge leads to transverse current at the off-current state and also a transverse potential between the upper and lower edges of ZGNR.    

This paper is organized in the following sections: After a comprehensive introduction, Hamiltonian and formalism will be presented in the subsequent section. The current-voltage characteristic curves and the physics behind of NDR, self-consistent charge and potential profiles and also the interplay between strain and charging effects is investigated in the result's section. Furthermore, spatial profiles of local bond currents are also studied at the on and off-current state. At the end we conclude all achievements in the conclusion section.   

\section{Hamiltonian and Formalism}
A schematic view of a field-effect transistor based on zigzag graphene nanoribbons is represented in Fig. \ref{fig:intro} such that one can apply tensile or compressive strain on ZGNR with a given strength and direction. The e-e interaction is considered to be long-range accompanied with an onsite Hubbard term simulating with an electrostatic Green's function which is calculated through the solution of the Poisson equation inside the scattering region. Dirichlet boundary condition would be located on the surface of semi-infinite electrodes. Therefore Hamiltonian of the scattering portion is described as\cite{cheraghchi2008negative}:
\begin{equation}
\begin{split}
& H=\sum_{i=1}^{N} \left[\varepsilon_{i} + u_{i}^{ext} +\sum_{j=1}^{N} V_{ij} \delta n_{j}\right] C^{\dagger}_{i}C_{i} 
\\
& + \sum_{<ij>} t_{ij} \left( C^{\dagger}_{i}C_{j} +C_{i}C^{\dagger}_{j}\right)
\end{split} 
\label{eq:1}
\end{equation}
Where, $ \varepsilon_{i} $ is the on-site energy of the $i^{th}$ site which is able to vary by the gate voltage ($V_g$). Laplace solution of the electrostatic potential ($ u_{i}^{ext} $), results in a linearly distributed potential profile along the scattering portion ($ \left[ (x_{i}-x_{L})/L-1/2 \right]V_{SD}  $) with the given boundary condition ($ u_{i}^{ext}(x_{L/R})=\pm V_{SD}/2 $) ($L$ is the length of the scattering region and $ V_{SD} $ is the source-drain bias voltage). 
The third term refers to the direct coulomb interaction representing electrostatic potential at site $i^{th}$ which is related to the electrostatic potential caused by variation in the self-consistent charge density $ \delta n_{j} $ at site $j^{th}$ relative to its initial value due to source-drain voltage. One of the proposals for the electrostatic green's function with cylindrical symmetry is considering as the following\cite{jackson}:
 \begin{equation}
 V(\overrightarrow{r},\overrightarrow{r\prime})=2\int^{\infty}_{0} dk \jmath_{0}(\alpha k)
 \dfrac{sinh(kx_{<})sinh(k(L-x_{>}))}{sinh(kL)}
 \label{eq:2}
 \end{equation} 
 \begin{equation}
\alpha=\sqrt{(x-x\prime)^2+(z-z\prime)^2+U_{H}^{-2}}
\label{eq:3}
\end{equation}  
 where $U_H$ is the Hubbard parameter which is about $4 t_0$ ($t_0$$\approx 2.7eV $ is the hopping energy between carbon bonds). $ \jmath_{0} $ is zeroth Bessel function. This electrostatic potential is modeled by the Ohno-Klopmann (OK)\cite{ohno1964some,klopman1964semiempirical,esfarjani1998self} kernel in which on-site potential is attributed to $U_H$. In this electrostatic potential, image charges which are induced in electrodes, are taken into account in our formalism\cite{cheraghchi2008negative}. Image charges break molecular symmetries in molecular junctions giving rise to NDR in the I-V curve\cite{Kaasbjerg2011}. \\
 
The last term $ t_{ij} $ is the kinetic energy which is affected by strain in both; central portion as well as electrodes. Each carbon atom has three nearest neighbor hopping representing as $t_{1}=t(\Delta_{1}),t_{2}=t(\Delta_{2}),t_{3}=t(\Delta_{3}) $ where $ \Delta_{i} $ are the connection vectors to nearest neighbour atoms. These connection vectors are varied under application of strain by using the following transformation $\Delta=(I+\epsilon)\Delta_{0} $ where $\Delta_0$ is equilibrium vectors \cite{pereira2009tight} and $ I $ is a unitary matrix.  $\epsilon$ is the stress tensor defining as the following:
\begin{equation}
\epsilon=\varepsilon\left( 
\begin{array}{cc}
cos^{2}\theta-\nu sin^{2}\theta & (1+\nu) cos\theta sin\theta \\ 
(1+\nu) cos\theta sin\theta & sin^{2}\theta-\nu cos^{2}\theta
\end{array} 
\right)  
\end{equation}
where $\varepsilon$,  $ \nu=0.165 $, $\theta$ are the strain strength, Poisson's ratio and the strain direction in respect to the ribbon axis, respectively. For the tensile/compressive strain, $\varepsilon$ is greater/lesser than zero. An exponential variation is one of the popular assumptions for the modification of hopping parameter caused by variation in the bond length under strain \cite{pereira2009tight}. So it is assumed that
\begin{equation}
t_{i}(\Delta_{i})=t_{0} e^{-3.37(\dfrac{\Delta_{i}}{a_{0}}-1)}
\end{equation}
where $ a_ {0}=1.42 \AA$ is the equilibrium bond length. The modified connection vectors ($\Delta_i$) for some strain directions are given in Tab. \ref{table:Tab1}.  In this paper, all energies are scaled to $t_0$. 
\\
Hamiltonian depends on the electron density. Therefore, we use self-consistent non-equilibrium green's function formalism (NEGF) to calculate charge density and converged Hamiltonian \cite{cheraghchi2010gate}. Charge density is departed to the equilibrium and non-equilibrium parts ($n=n^{non-eq}+n^{eq}$) which is defined in terms of the green's function as the following:
\begin{equation}
n_{i}^{eq}=-1/\pi\int_{-\infty}^{E_{F}-V_{SD}/2} Im [G_{ii}^{r}(E)]dE 
\end{equation}
\begin{equation}
\begin{split}
& n_{i}^{non-eq}=1/2\pi\int_{E_{F}-V_{SD}/2}^{E_{F}+V_{SD}/2} [G^{r}(\Gamma_{L}f_{L}+\Gamma_{R}f_{R})G^{a}]_{ii}dE
\\ 
\end{split}
\end{equation}
where $G^{r,a}(E,V)=\left[ (E \pm i0^{+})-H-\Sigma_{L}^{r,a}-\Sigma_{R}^{r,a}\right]^{-1}$ is the retarded and advanced green's function which is related to the retarded and advanced self-energies coming from the left and right electrodes ($\Sigma_{L,R}^{r,a}$). Here, $\Gamma_{R,L} $ is the escaping rates of electrons to the electrodes and $f_{R,L} $ is the Fermi function of electrodes assuming to be at zero temperature. $E_F$ is the Fermi energy of the system which is consider to be fixed at $E_F=0$.\\ 
Although the direct Coulomb interaction is present in our calculation, transport through nanoribbons is still coherent. Therefore it is allowed to use the Landauer formula for the current at zero temperature whenever converged Hamiltonian is achieved\cite{nardelli1999electronic,liang2004electrostatic}.
\begin{equation}
I(V)= \dfrac{2e}{h} \int_{\mu_{L}}^{\mu_{R}} T(E,V) dE
\label{eq:}
\end{equation}
where $ \mu_{L/R} $ is the chemical potential of the left and right electrodes. $T(E,V )$ is the bias dependent transmission coefficient.
\begin{equation}
T(E,V)=Tr[\Gamma _{L}G^{r}\Gamma _{R}G^{a}]
\label{•}
\end{equation}

\begin{table}
\includegraphics[width=0.45\linewidth]{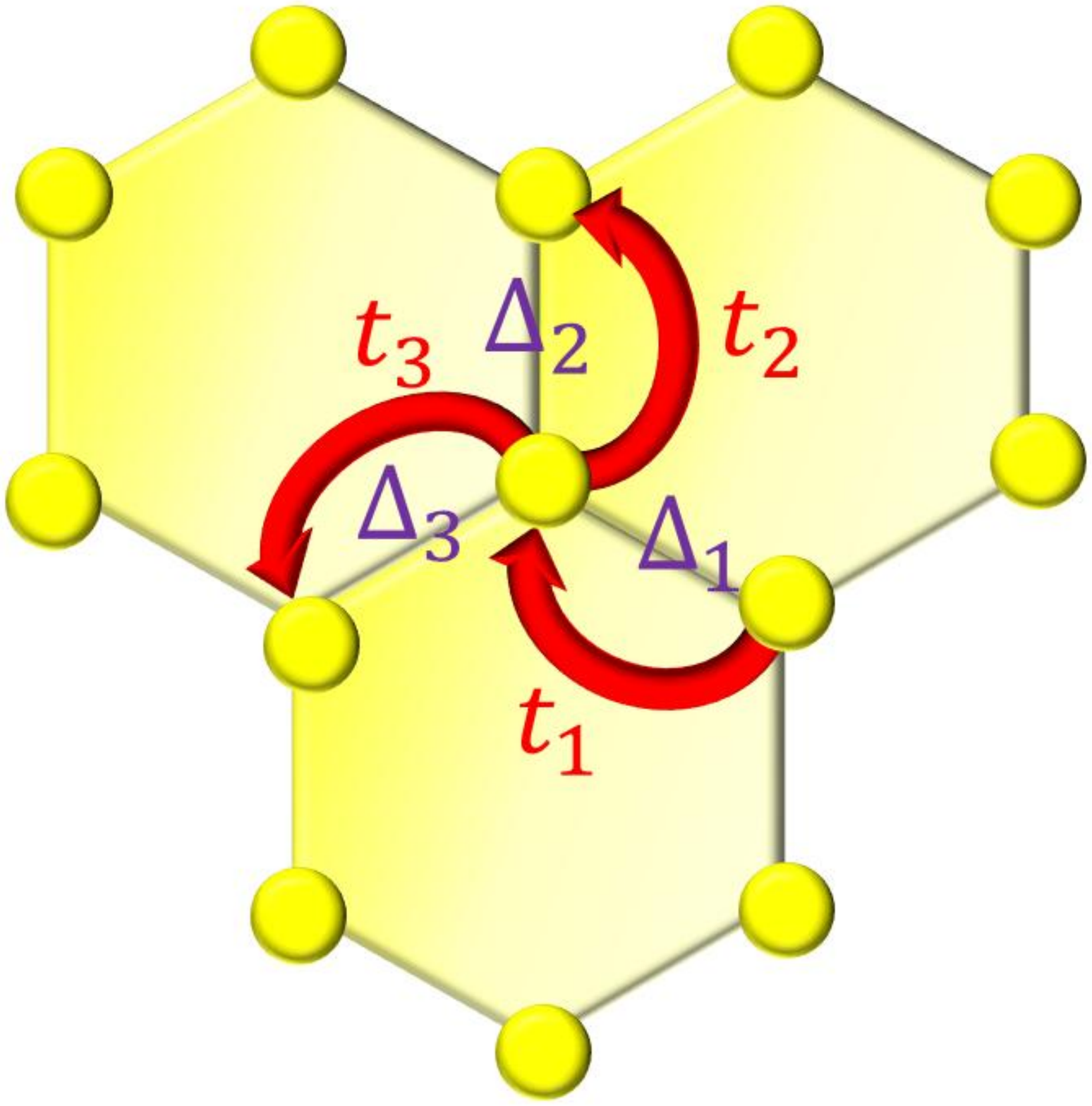}
\begin{tabular}{|c  c  c|}
\hline
\diagbox[dir=SE,width=3.7em] {$ \theta $} {$ \varepsilon > 0 $}  & Bond Length & Hopping Parameter \\
\hline
\hline
$ 0  $ & $ \Delta_{2}< \Delta_{3}=\Delta_{1} $& $ t_{2}>t_{1}=t_{3} $ \\
\hline
$ \pi/6 $ & $ \Delta_{3}>\Delta_{2}=\Delta_{1} $& $t_{3}<t_{2}=t_{1}  $\\
\hline
$ \pi/4 $ & $ \Delta_{3}>\Delta_{2}>\Delta_{1} $& $t_{3}<t_{2}<t_{1}  $ \\
\hline
$ \pi/3 $ & $ \Delta_{3}=\Delta_{2}>\Delta_{1} $& $ t_{3}=t_{2} < t_{1} $ \\
\hline
$ \pi/2 $ &  $ \Delta_{2} > \Delta_{3}=\Delta_{1} $ & $ t_{2} < t_{3}=t_{1} $\\
\hline
\end{tabular}
\\ 
\begin{tabular}{|c  c  c|}
\hline
\diagbox[dir=SE,width=3.7em]{$\theta$}{$\varepsilon < 0$}  & Bond Length & Hopping Parameter \\
\hline
\hline
$ 0  $ & $ \Delta_{2}> \Delta_{3}=\Delta_{1} $& $ t_{2}<t_{1}=t_{3} $ \\
\hline
$ \pi/6 $ & $ \Delta_{3}<\Delta_{2}=\Delta_{1} $& $t_{3}>t_{2}=t_{1}  $\\
\hline
$ \pi/4 $ & $ \Delta_{3}<\Delta_{2}<\Delta_{1} $& $t_{3}>t_{2}>t_{1}  $ \\
\hline
$ \pi/3 $ & $ \Delta_{3}=\Delta_{2}<\Delta_{1} $& $ t_{3}=t_{2} > t_{1} $ \\
\hline
$ \pi/2 $ &  $ \Delta_{2} < \Delta_{3}=\Delta_{1} $ & $ t_{2} > t_{3}=t_{1} $\\
\hline
\end{tabular}
\caption{Comparison of the bond lengths and hopping energies of the three nearest neighbour atoms for: (top table) the tensile strain ($\varepsilon >0$), (bottom table) the compressive strain ($\varepsilon <0$) along different strain directions.}
\label{table:Tab1}
\end{table}
\section{Results}
\subsection{I-V characteristic curves under uniaxial strain}
Observation of negative differential resistance (NDR) phenomena in the current-voltage characteristic curves gives us hopes to be able to manipulate electronic switch devices based on straintronics. An interesting category of such NDR's is originated from the quantum transport gaps instead of the energy gap which Esaki diodes\cite{Esaki} are based on. Quantum transition rules in ZGNR's with even number of zigzag chains could results in NDR in I-V curve. In the following, we investigate the effect of tensile and compressive uniaxial strain in different directions and strengths on the I-V characteristic curve of ZGNRs in the presence and absence of electron-electron interaction.\\ 

\begin{figure}[h]
\centering
\includegraphics[width=0.75\linewidth] {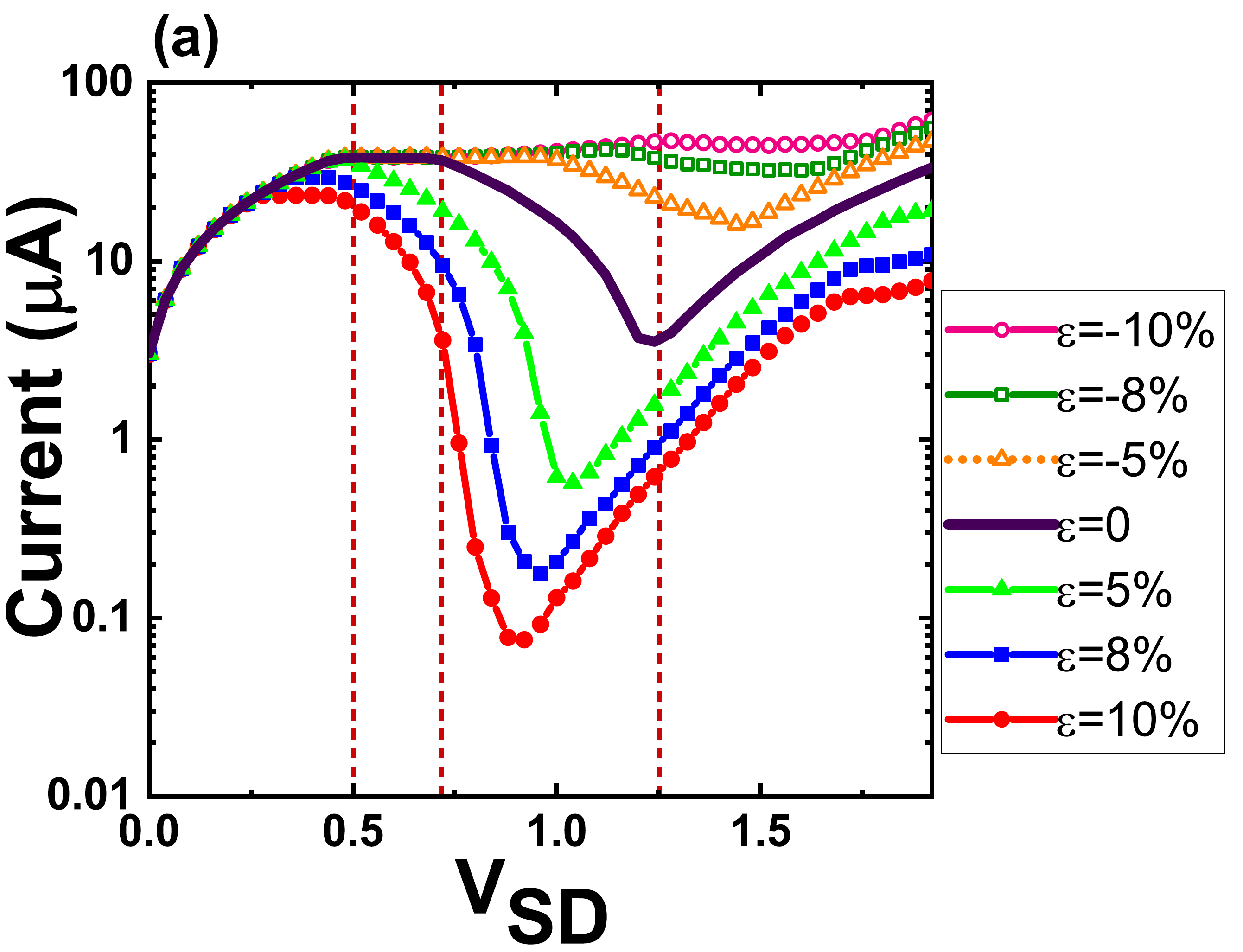}
\includegraphics[width=0.75\linewidth] {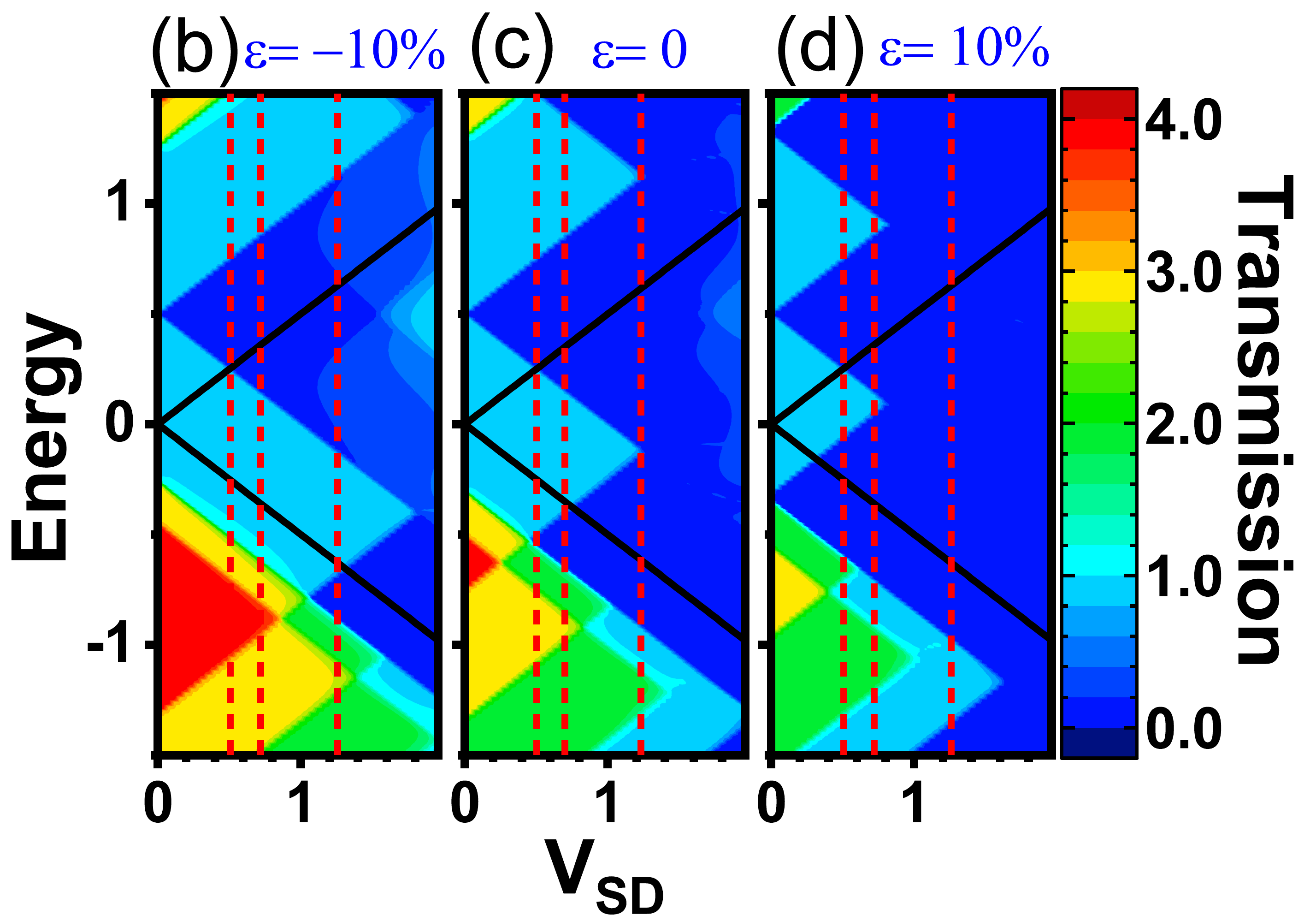}
\caption{Current-voltage(I-V) characteristic curve of Zigzag graphene nanoribbon ZGNR(M,N)=(4,6) under application of uniaxial strain in a) $\theta =0 $ direction. The gate voltage is considered to be as ($ V_{g}=0.5$). b) a 3D contour-plot of transmission in terms of energy and source-drain voltage for strain at $\theta =0 $ direction for three values of strain strength $\varepsilon$=-10 $\%$,0,+10$\%$.  Red Dashed lines correspond to bias voltages shown in I-V curves. Here ZGNR is considered to be non-interacting.}
\label{fig:current0-90}
\end{figure} 
Let us first investigate the effect of tensile and compressive strain along the $\theta=0$ direction on I-V curve as shown in Fig.\ref{fig:current0-90} when e-e interaction is not present. As readily seen, $ I_{on}/I_{off} $ ratio increases by increasing of tensile strain from 5\% to 10\%. However, on/off ratio of current decreases for the compressive strain so that NDR is being disappeared by increasing strain up to -10\%. 

To understand the physics behind of this behavior let us look at the Tab. \ref{table:Tab1}. Along the $\theta=0$ direction, tensile/compressive strain causes to decrease/increase hopping energy and so as a result, it is reasonable to have narrower/wider band structure (See more in Appendix \ref{appendix:a}). Considering the Poisson ratio, the bond lengths and also hopping energies of the given site with its three nearest neighbor atoms are compared with each other in Tab. \ref{table:Tab1} for different strain directions. As seen of the table, hopping energies along the nanoribbon axis is decreased (increased) for the tensile strain along $\theta=0 (\pi/2)$ direction. So the band spectrum would be narrower (wider) than the un-strained case. On the other hand, in some energy ranges, transport channels will be decreased/increased when tensile/compressive strain is applied (Appendix \ref{appendix:a}). As a conclusion of such narrow/wide band spectrum, as shown in 3D contour-plot of transmission in terms of energy and $ V_{SD} $, transmission in the integration window (dark bold lines in Fig. \ref{fig:current0-90} b-d) is decreased/increased (Fig. \ref{fig:current0-90} d/b). 

In fact, NDR is originated from two blocked transport channels; one of them is caused by the parity selection rule, the other one is caused by the blocked transition between disconnected bands\cite{cheraghchi2010gate}. These blocked transport channels survive under application of longitudinal and transverse strain. The details of this claim is demonstrated in Appendix \ref{appendix:a}. 

It was also checked that all feature of I-V curve will be remained unchanged for the case of uniaxial strain along the $\theta=\pi/2$ direction except that I-V curves of tensile strain is replaced by compressive one in Fig.\ref{fig:current0-90}a. So it is feasible to manipulate a nano-electromechanical switch by replacing tensile/compressive strain along $\theta=0$  to compressive/tensile strain along $\theta=\pi/2$ . In other words, by switching between two perpendicular strain and also type of stress (tensile or compressive strain), current passing through nanoribbon can be switched in its off or on-state.  \\

\begin{figure}[h]
\centering
\includegraphics[width=0.7\linewidth] {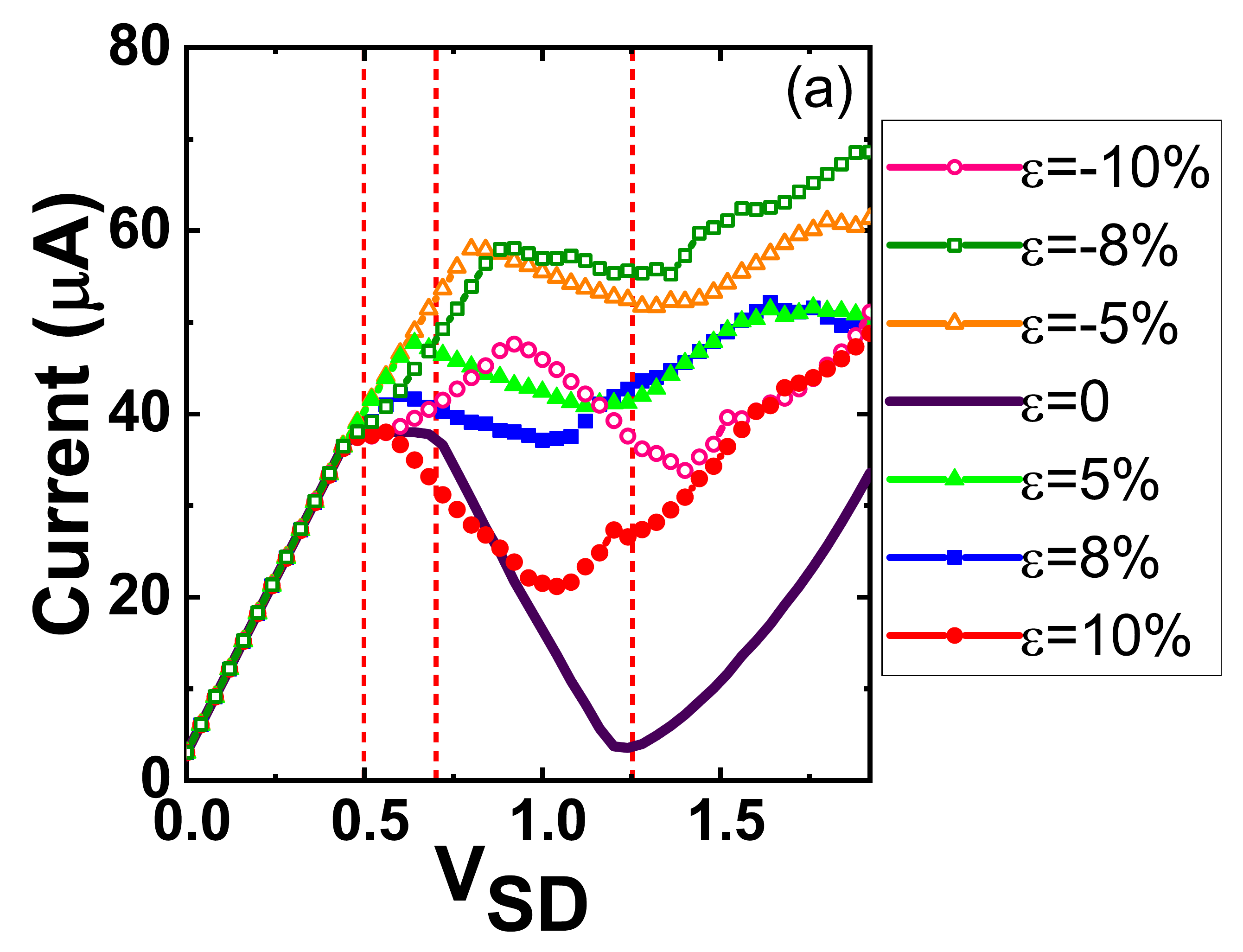}
\includegraphics[width=0.7\linewidth] {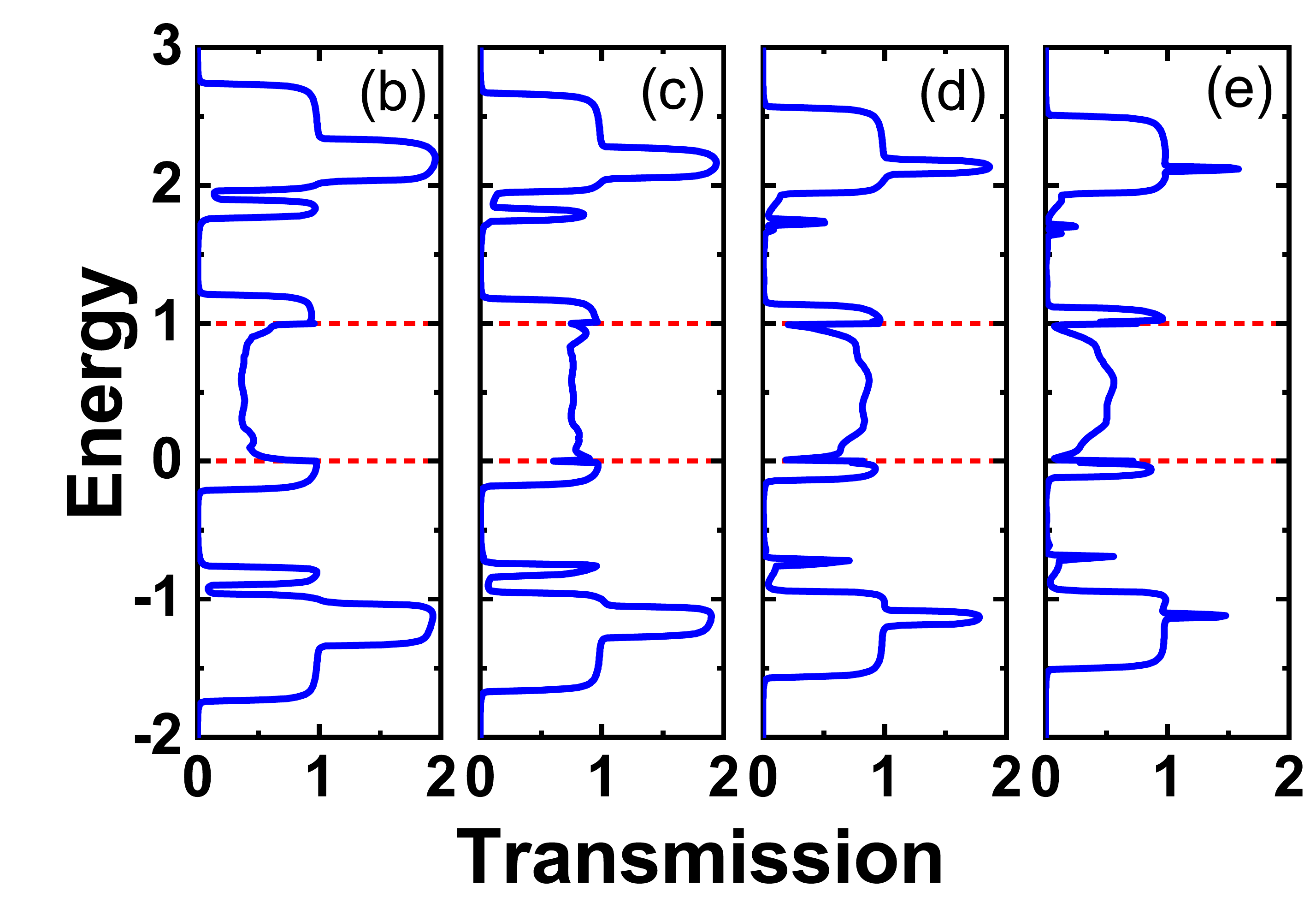}
\includegraphics[width=0.7\linewidth] {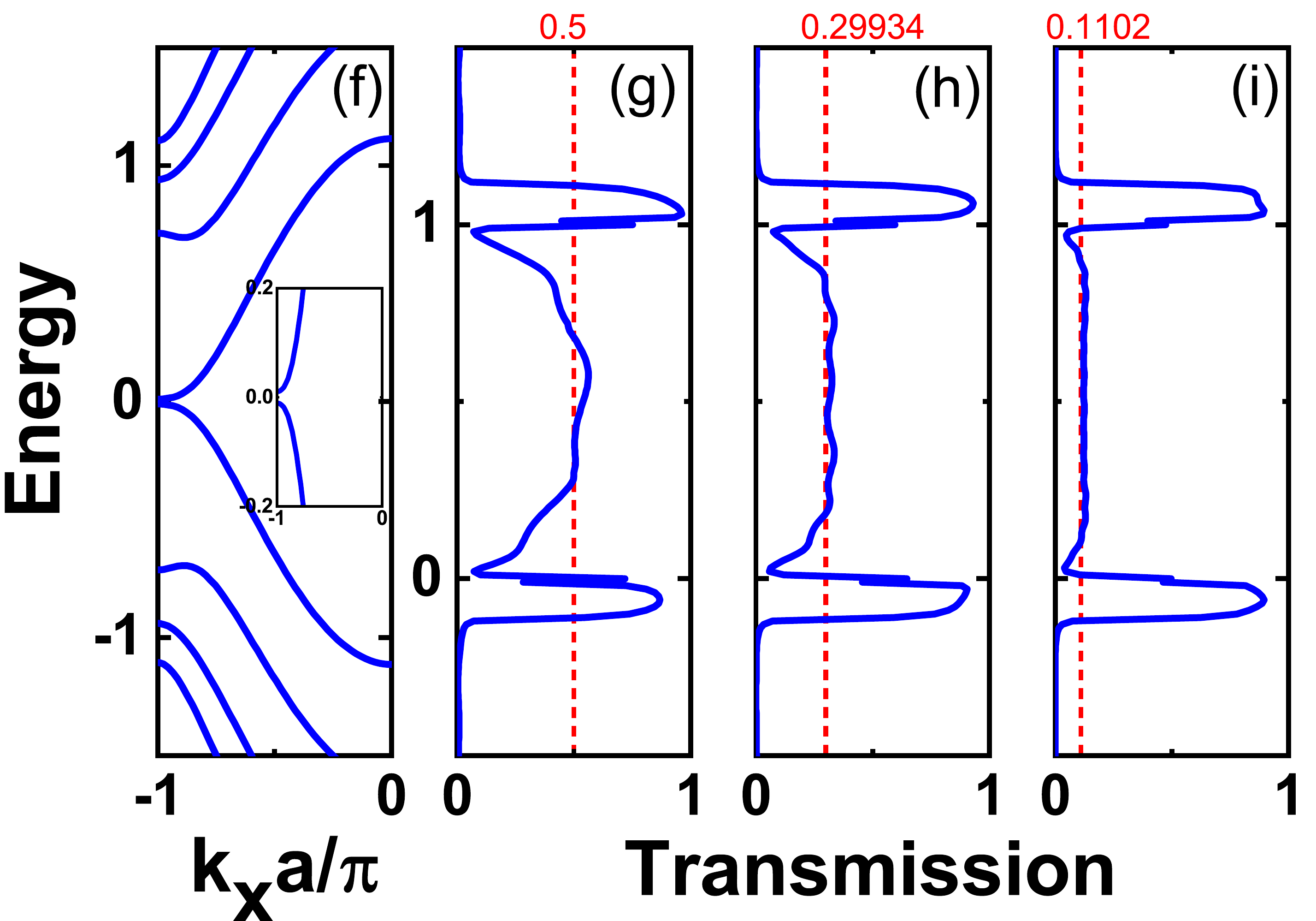}
\caption{a)Current-voltage(I-V) characteristic curve of ZGNR(M,N)=(4,6) under application of the gate voltage ($ V_{g}=0.5$) for different strain strength applied along the $\theta =\pi/4$ direction. The second panel contains transmission for different strain strengths: b) $ \varepsilon=5\% $, c) $ \varepsilon= 8\% $, d) $ \varepsilon= 10\%$, e) $ \varepsilon= 12\%$. Here, $V_{SD}=1 $. Red dash lines show the limits of central band where parity conservation is violated. f)Energy spectrum after application strain ($ \theta=\pi/4 ,\varepsilon=10\% $). The inset figure is magnification of the central band of part (f) that shows the splitting of central energy bands at the Brillouin zone boundary. The comparison of transmission coefficients in the presence of strain $ \theta=\pi/4, \varepsilon=10\%$ for ZGNRs with 4 zigzag chains in width and g)6, h)12, i)24 unit cells in length. }
\label{fig:oblique_strain}
\end{figure} 

For the oblique uniaxial strains, for example, strain along the direction of $ \theta=\pi/4 $, $ I_{on}/I_{off} $ ratio drastically decreases in compared to the un-strained case for both of tensile and compressive strain. In this strain direction, the geometry symmetry is broken. So the parity conversation in even ZGNR's is no longer valid and as a consequence transport gap coming from this transition rule would be filled. The second transition rule, namely, forbidden transition between disconnected bands, is still valid. In fact, disappearing of NDR refers to the filling of the transport gap representing in Fig. \ref{fig:oblique_strain}(b-e). For the sake of clarification, let us look at transmission coefficient in different strain strengths. As demonstrated in Fig. \ref{fig:oblique_strain}(b-e), up to 8\% strain along $\pi/4$ direction, the transport gap is being filled and after this critical strain, transmission coefficient inside the gap is again decreasing. The first part of phenomena is related to the breaking of geometry symmetry and consequently violation of the parity conservation rule. However, decreasing of transmission comes back to separating of the central bands inducing by tensile strain stronger than 10\%. This separation is displayed in Fig. \ref{fig:oblique_strain}f. To check that this decreased transmission refers to disconnected central bands, transmission at the band center is investigated by variation in the nanoribbon length as shown in Fig. \ref{fig:oblique_strain}(g-i). An increased length concludes more decrease in transmission coefficient at the band center. So having long ribbons causes to have smooth variation of the applied bias along the nanoribbon such that electrons are scattered between those band states which belong to continuous bands\cite{Grosso1,Grosso2}. However, turning e-e interaction on concludes interesting and non-trivial results which can affect non-interacting transport properties.          

\subsection{Charging effect}
\begin{figure}[h]
\centering
\includegraphics[width=0.85\linewidth] {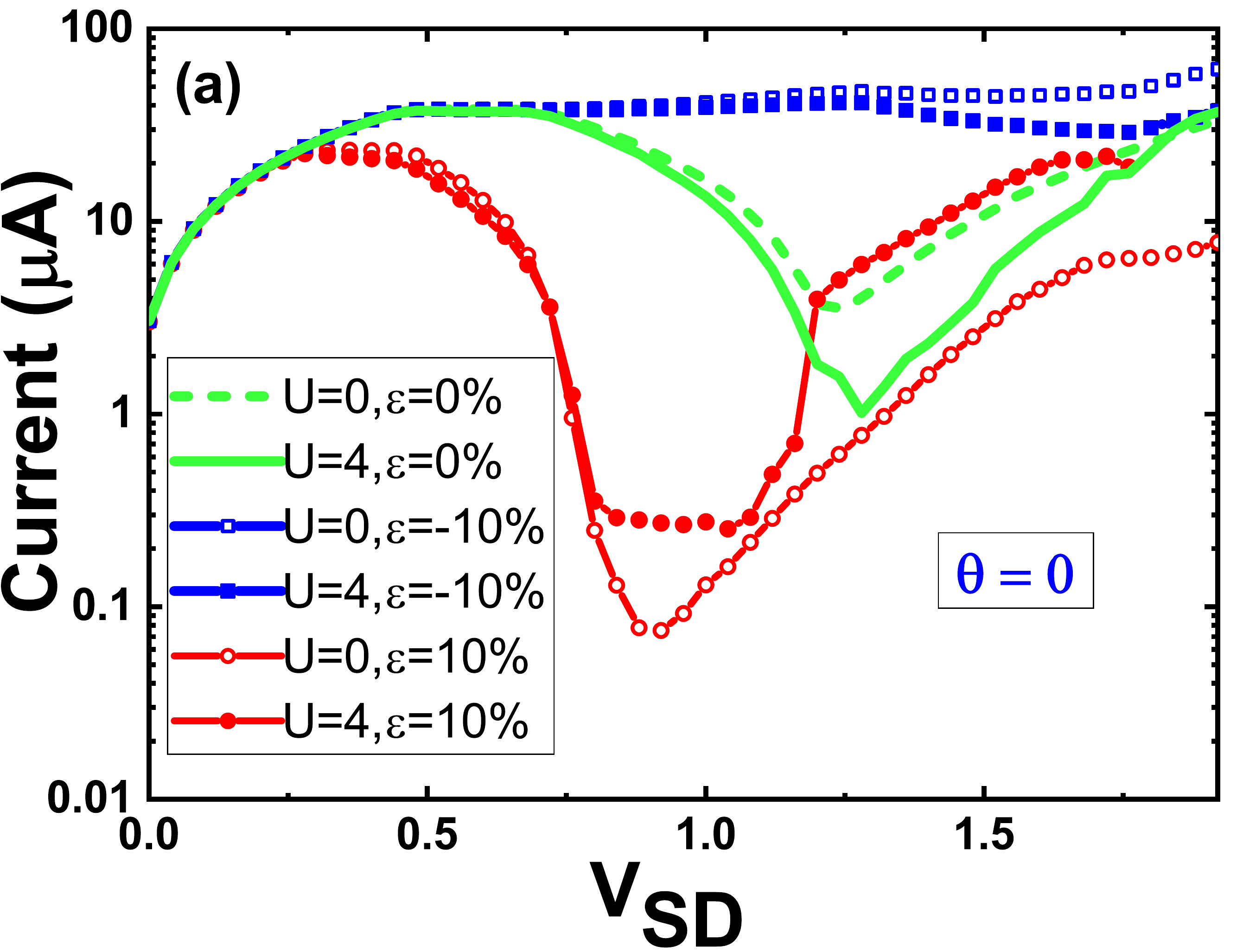}
\includegraphics[width=0.85\linewidth] {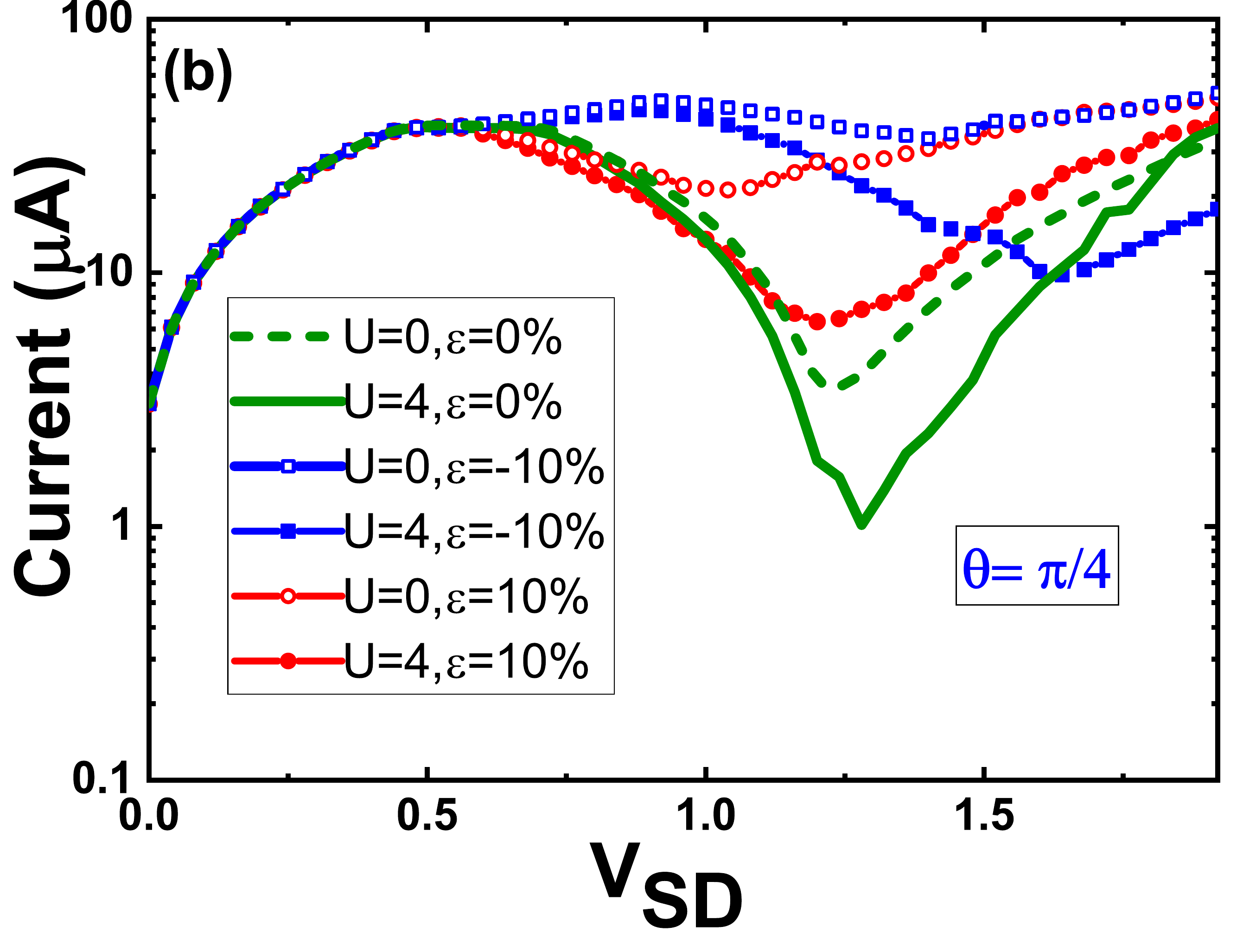}
\caption{Current-voltage characteristic curve of ZGNR(M,N)=(4,6) under application of different strain values for ($ V_{g} =0.5$) a)$ \theta =0 $ and b)$ \theta =\pi/4 $ in the presence of electron-electron interaction. Here U is the Hubbard parameter.}
\label{fig:currentU}
\end{figure}

\begin{figure}[h]
\centering
\includegraphics[width=0.95\linewidth] {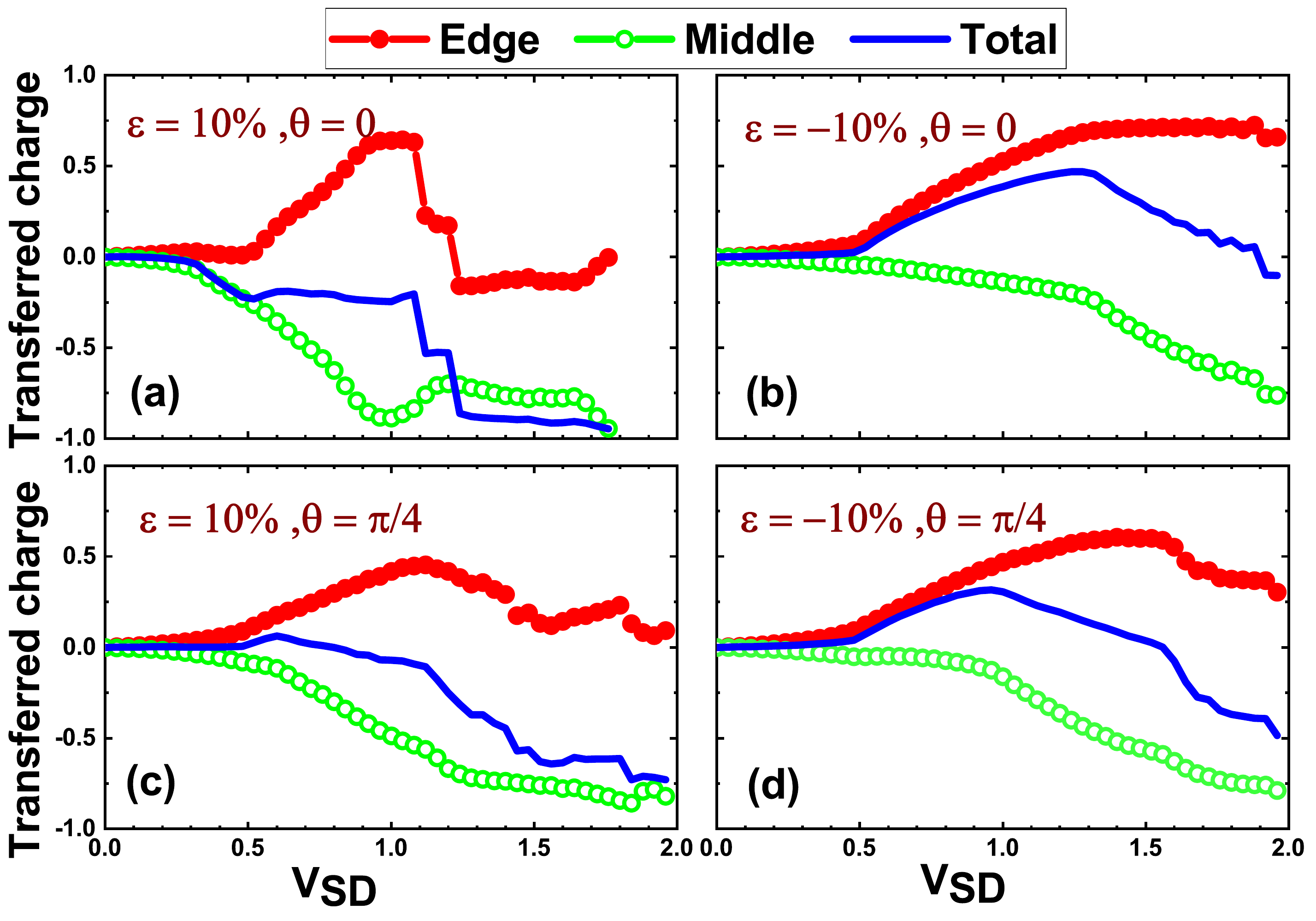}
\caption{Edge,middle and total transferred charge in terms of applied source-drain voltage $ V_{SD} $ for ZGNR(4,6),$ V_{g}=0.5$ in different strains: a)$ \theta=0 ,\varepsilon=10\% $, b)$ \theta=0 ,\varepsilon=-10\% $, c)$ \theta=\pi/4 ,\varepsilon=10\% $, d)$ \theta=\pi/4 ,\varepsilon=-10\% $}
\label{fig:charge}
\end{figure}

The interplay between long range Coulomb interaction and strain is represented in the I-V characteristic curve shown in Fig. \ref{fig:currentU}. In the absence of strain, turning the e-e interaction on causes to decrease off-current ($I_{off}$). The reason comes back to those conducting channels in which transmission coefficient would be suppressed because of electron scattering induced by the electrostatic potential profile producing by the transferred charge between the source-drain electrodes and central portion\cite{cheraghchi2010gate}. The decrease of off-current in the presence of e-e interaction is also observed in the strain (compressive) applied along the $\theta=0$ direction (Fig. \ref{fig:currentU} a) and the  oblique strain such as $\theta =\pi/4$ (Fig. \ref{fig:currentU} b). However, when ZGNR is affected by the tensile strain along the $\theta=0$ direction, the off-current shows an increase behavior in compared with the non-interacting system (Fig. \ref{fig:currentU} a). For searching of the physics behind of these behaviors, it is useful to investigate self-consistent electrostatic charge and potential profiles under applied strain. 
Total charge variation is almost zero for the voltages smaller than $ V_{on} $(threshold voltage for emerging NDR which corresponds to on-current) as represented in Fig. \ref{fig:charge}. However, in the voltage range $[V_{on},V_{off}]$, depending on the strain type, total charge is depleted or accumulated in the central portion. Fig. \ref{fig:charge}(a,c) demonstrates that in the mentioned voltage range, total charge is depleted ($\delta n<0$) for the tensile strain while it is accumulated ($\delta n>0$) on the central portion for the compressive strain as shown in Fig. \ref{fig:charge}(b,d). This phenomenon is independent of the strain direction. In more details, charge distribution along the system is categorized by the edge and middle transferred charge in Fig. \ref{fig:charge} which departs charge accumulation on the edges of ZGNRs from the charge depletion of the middle part of ZGNRs. The dominant phenomena for all cases would be charge accumulation on the edges of ZGNR which affects spatial profile of local bond currents and as consequence, I-V characteristic curve.

\begin{figure}[h]
\centering
\includegraphics[width=1\linewidth] {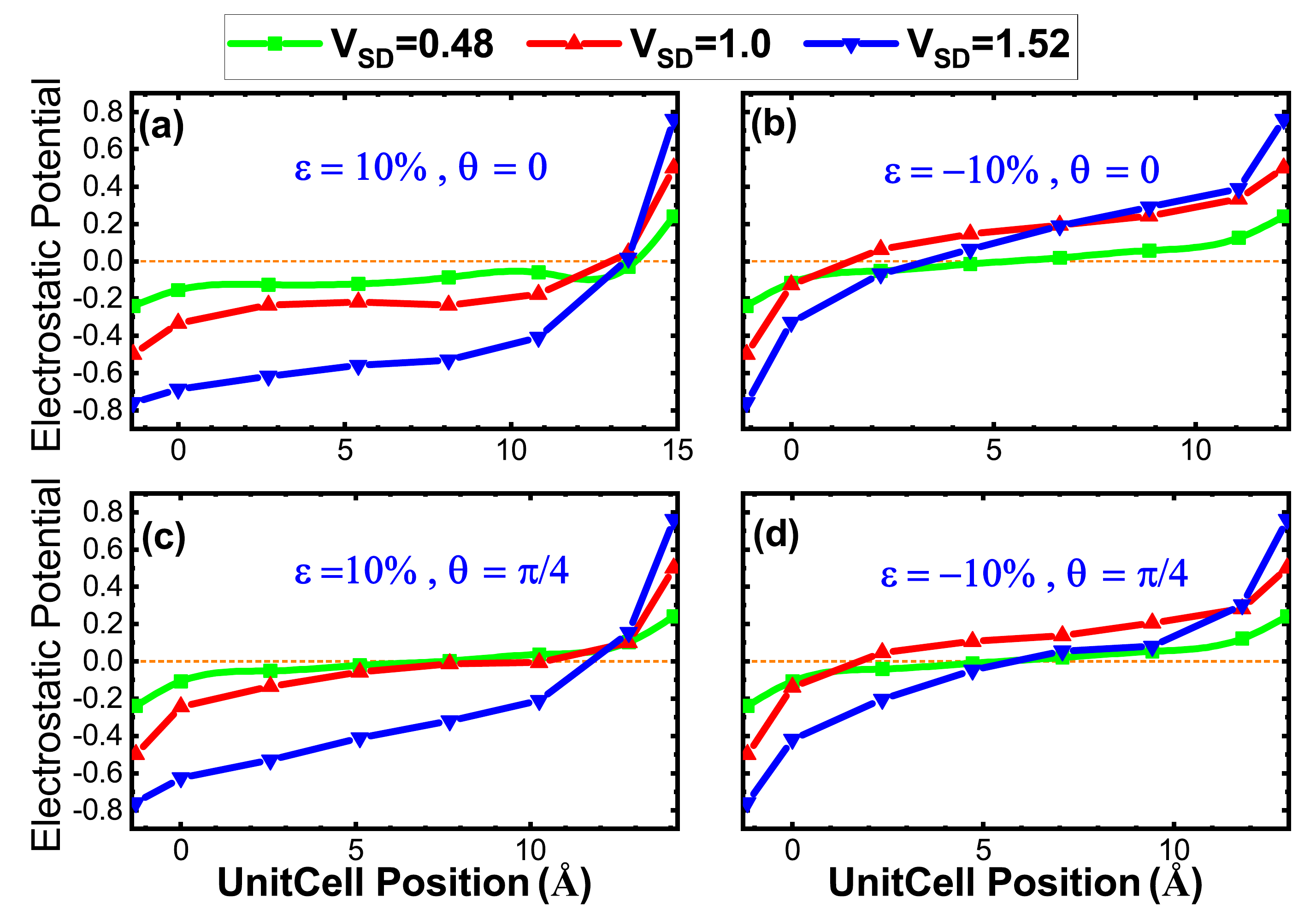}
\caption{Electrostatic potential per unit cell in terms of X-component of unit cell positions in three voltages $ V_{SD}= $0.48 ,1 ,1.52  under different stain directions and values: a)$ \theta=0$, $\varepsilon=10\% $ b)$ \theta=0$, $\varepsilon=-10\% $ c)$ \theta=\pi/4$, $\varepsilon=10\% $ d)$ \theta=\pi/4$, $\varepsilon=-10\% $.}
\label{fig:xpot}
\end{figure}

\begin{figure}[h]
\centering
\includegraphics[width=1\linewidth] {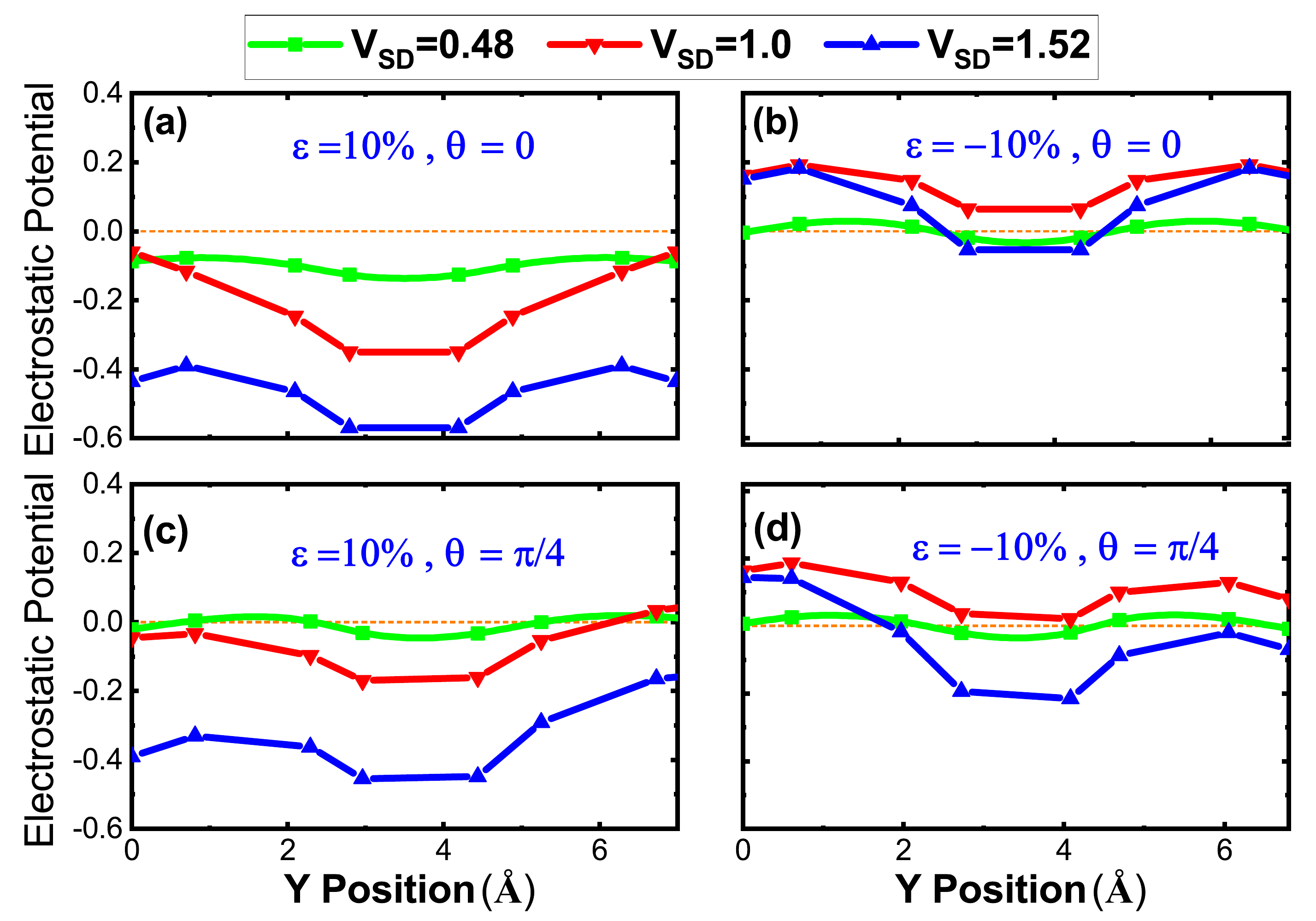}
\caption{Electrostatic potential per zigzag chain in terms of Y-component of atom position in three voltages $V_{SD}$ = 0.48, 1, 1.52 for different stain directions and strengths: a)$ \theta=0$, $\varepsilon=10\% $ b)$ \theta=0$, $\varepsilon=-10\% $ c)$ \theta=\pi/4$, $\varepsilon=10\% $ d)$ \theta=\pi/4$, $\varepsilon=-10\% $}
\label{fig:ypot}
\end{figure}

The longitudinal and transverse electrostatic potential profiles are shown in Figs. \ref{fig:xpot} and \ref{fig:ypot}.
The average of electrostatic potential over each unit cell of ZGNR is displayed in Fig. \ref{fig:xpot} while the longitudinal average along the axis of nanoribbon is drawn in Fig. \ref{fig:ypot}. Independent of the strain direction, as seen in Fig. \ref{fig:xpot} (a,c), the dominant behavior in the tensile strain ($\varepsilon=10 \%$), would be a sharp potential drop asymmetrically on the source side and a flat potential among the bulk of ZGNR. On the other hand, the external potential is fully screened by electron depletion in the source side. This sharp drop of the electrostatic potential results to large variation of electron momentum and causes to fail the transition rules which are the origin of transport gaps\cite{cheraghchi2010gate,Grosso1,Grosso2}. It should be reminded that for the oblique strains, in the non-interacting case, the parity selection rule had been already failed because of geometry symmetry breaking. As it will be presented later, in the oblique strains, imbalance accumulation of the edge charge on the upper and lower edges, could be one of the reasons for decreasing off-current in I-V curve. Furthermore, an enhancement of the off-current in the tensile strain along $\theta=0$ direction comes back to breaking of those transition rules which result in transport gaps (Fig. \ref{fig:currentU} a).

In the compressive strain ($\varepsilon=-10 \%$)(Fig. \ref{fig:xpot} (b,d)), potential drop happens at the contacts. On the other words, external bias is screened by the central portion symmetrically such that the potential far from the contacts is almost smooth. In this case, the feature of the potential profile is also independent of the strain direction. As it will be mentioned later, the only difference would be a transverse voltage which appears on the edges of ZGNR when an oblique strain is applied. In agreement with charge depletion in the tensile strains shown in Fig. \ref{fig:charge}, electrostatic potential would be at lower level in compared to the compressive strain as depicted in Fig. \ref{fig:xpot}(a,c).

The potential well caused by the tensile strain is obvious if one looks at the transverse potential profile shown in Fig. \ref{fig:ypot} (a-d). Moreover, the cavity in the transverse potential profile indicates that discharging happens mostly from the middle part of ZGNR. However, novel phenomena would emerge when the strain direction is oblique such as the direction of $\theta =\pi/4$. In this case, the transverse potential is anti-symmetric across the nanoribbon and as a result, a transverse voltage would be measurable in ZGNR. This anti-symmetric potential leads to a charge imbalance between the upper and lower edges. It should be mentioned that as we will show later, the local bond currents through nanoribbon also would be anti-symmetric across ZGNR.

\begin{figure}[h]
\centering
\includegraphics[width=0.75\linewidth] {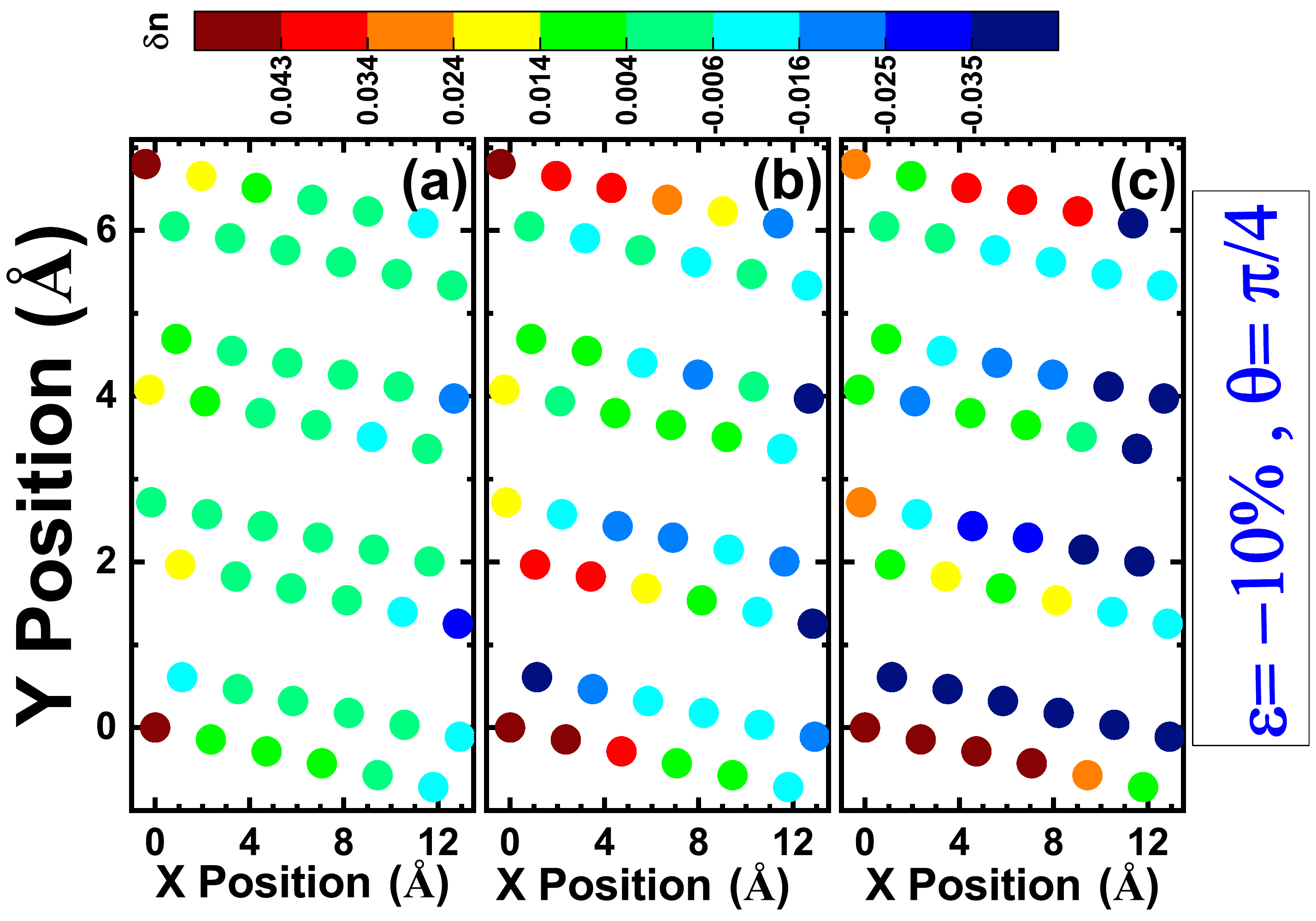}
\includegraphics[width=0.75\linewidth] {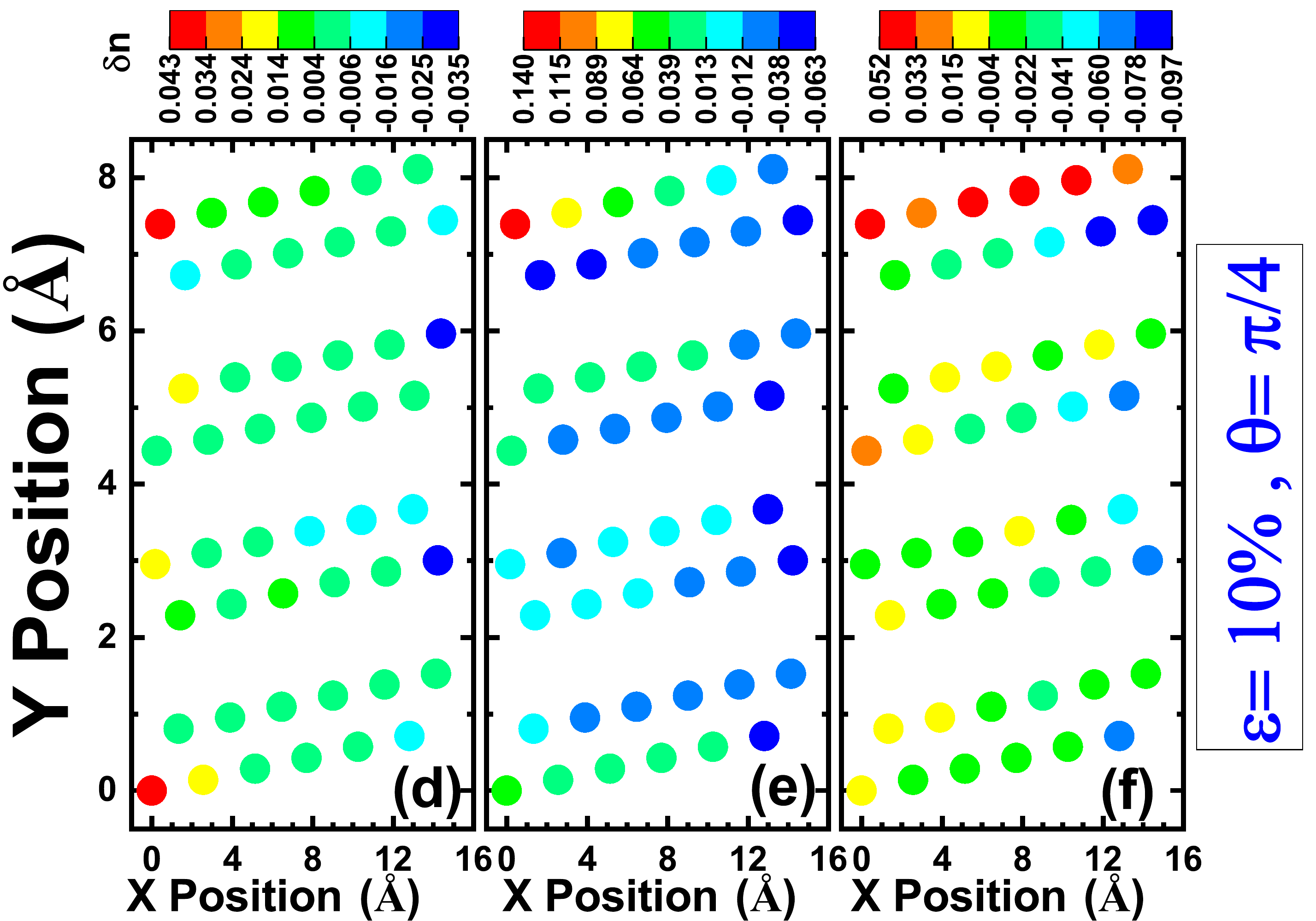}
\caption{Spatial profiles of transferred charge for the compressive (top panels $\varepsilon= -10\% $) and tensile (bottom panels $\varepsilon= +10\% $) strains along the $ \theta=\pi/4 $ direction for three source-drain applied biases $ V_{SD}=$   a,d) 0.48 , b,e) 1 , c,f) 1.52. Here the gate bias is $ V_{g}=0.5  $.}
\label{fig:compressivecharge}
\end{figure} 

The details of the above averaged charge and potential values are investigated at the spatial profiles of transferred charge separating by each atom for the oblique compressive and tensile strains applied along $\theta=\pi/4$ at three voltages of 0.48 and 1 and 1.52 which is shown in Fig. \ref{fig:compressivecharge}. At low bias voltages, as seen in Fig. \ref{fig:compressivecharge} a, charge is depleted from the source side and accumulated on the drain side of ZGNR. As the bias enhances, charge depletion would extend through the middle part of ZGNR and simultaneously, charge accumulation happens at the edges of ZGNR. Let us mention again that as it is clear, charge depletion for the tensile case (bottom panel; Fig.\ref{fig:compressivecharge} e,f) is more drastic than the compressive strain (top panel; Fig.\ref{fig:compressivecharge} b,c). Furthermore, as the bias increases, we observe that the charge accumulates on the lower edge more than on the upper edge for the compressive strain while it is vice verse for the tensile strain. This imbalanced charge accumulation at one edge in the tensile and compressive strains is due to the difference of the electrostatic potential at each edge as shown in Fig. \ref{fig:ypot}. Indeed, as seen in Tab. \ref{table:Tab1}, the applied strain along the $ \pi/4 $ direction breaks down the geometric symmetry of the system and the bond lengths and the hopping parameters change for the first neighbors of a carbon atom.

\begin{figure}[h]
\centering
\includegraphics[width=1\linewidth] {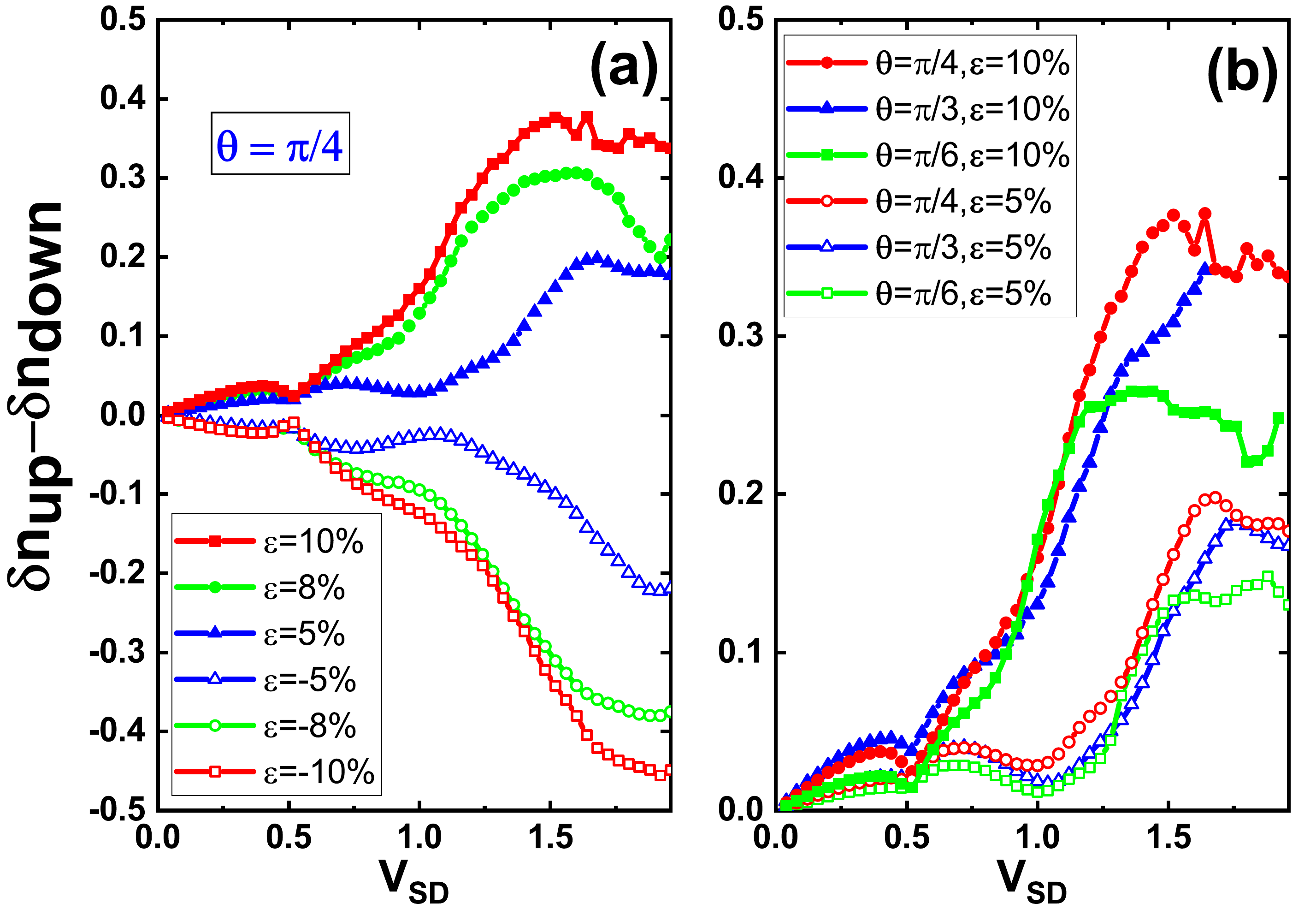}
\caption{Difference between the upper and lower edge transferred charge in terms of bias voltage under applied strain a) along the $ \theta=\pi/4 $ direction for different tensile and compressive strain strengths b) along different directions ($ \theta=\pi/6, \pi/4, \pi/3 $) and tensile strain strengths ($\varepsilon= 5 \%,10\%$).}
\label{fig:diffcharge}
\end{figure}

For the sake of completeness, we present difference of transferred charge density between the upper edge in compared to the lower edge of ZGNR in terms of the source-drain bias. Fig. \ref{fig:diffcharge}a demonstrates that for the tensile and compressive strains, by an enhancement in strength of strain, edge charge difference would increase. Although Fig. \ref{fig:diffcharge}a displays results for the oblique strain along the $ \theta =\pi/4 $ direction, the mentioned conclusion can be generalized for each direction of the applied strain. It is obvious that after on-current bias $V_{on}$ (for example here, $V_{SD}=0.5$), imbalance of accumulated charge between two edges increases by the source-drain bias. There is a minimum in the charge imbalance at the on-current state for all cases. So there is a relation between edge charge imbalance and longitudinal current. Fig. \ref{fig:diffcharge}b demonstrates that charge imbalance on the edges has its maximum value when strain is applied along the $\pi /4$ direction. However, on the charge imbalance, the strain strength is more effective than the strain direction.  
\subsection{Bond charge current}
The physics governing on charging effect is the charge conservation law which is manifested in the continuity equation giving rise to adjust charge accumulation or depletion on one hand, and the local non-equilibrium current flowing through ZGNR from the other hand. The charge continuity equation which is accessible by the Heisenberg equation is described as the following\cite{cheraghchi2011nonlinear,zarbo2007spatial}:
\begin{equation}
e\frac{dn^{non-eq}_{\bold i}}{dt}+\Sigma_{{\bold \Delta_j}}[J_{{\bold i},{\bold i}+{\bold \Delta_j}}-J_{{\bold i}+{\bold \Delta_j},{\bold i}}]=0
\label{continuity}
\end{equation}
where $J_{{\bold i},{\bold i}+{\bold \Delta_j}}$ is the charge current from site ${\bold i}^{th}$ to its nearest neighbour site ${\bold j}^{th}$. The local current induced by non-equilibrium charge is given by using the following equation,

\begin{equation}
J_{ij}=\frac{2e t_{ij}}{h}\int_{E_F-V_{SD}/2}^{E_F+V_{SD}/2} dE [G^{<}_{ij}(E)-G^{<}_{ji}(E)]
\end{equation}  

where lesser Green's function is defined as $$-iG^<=G^{r}(\Gamma_{L}f_{L}+\Gamma_{R}f_{R})G^{a}.$$ The resultant charge current which is coming out of site $i^{th}$, is determined as the following; ${\bold J}_{i}=\Sigma_{j} J_{ij}[\frac{\bold \Delta_j}{\Delta_j}]$. Here $j$ is the summation over nearest neighbor sites. It can be simply proved by Heisenberg equation that the formulation for the bond current is applicable for Hamiltonian presented in Eq. \ref{eq:1} in which diagonal terms depends on the charge density.  
 
\begin{figure}[]
\centering
\includegraphics[width=1\linewidth]{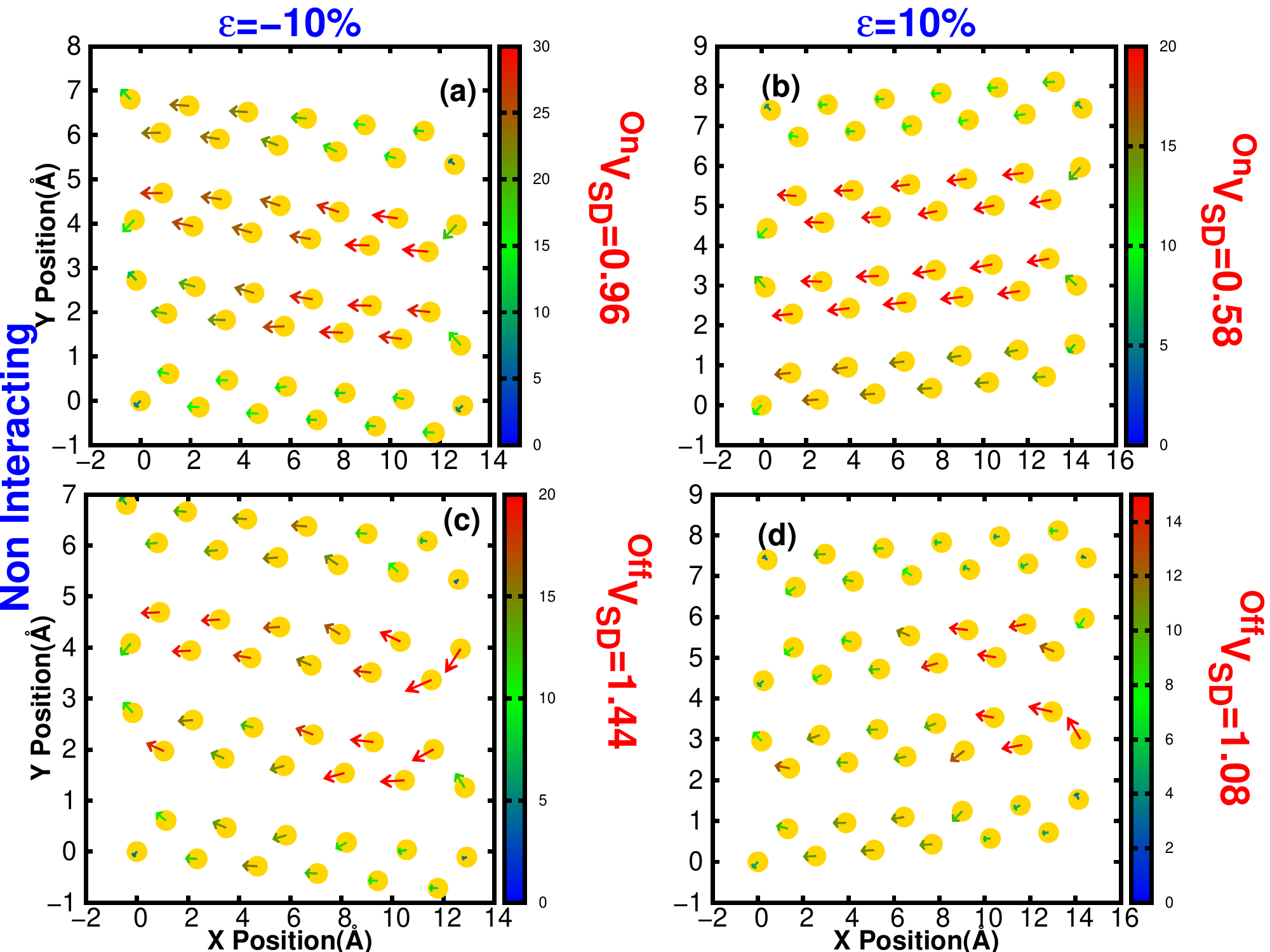} 
\includegraphics[width=1\linewidth]{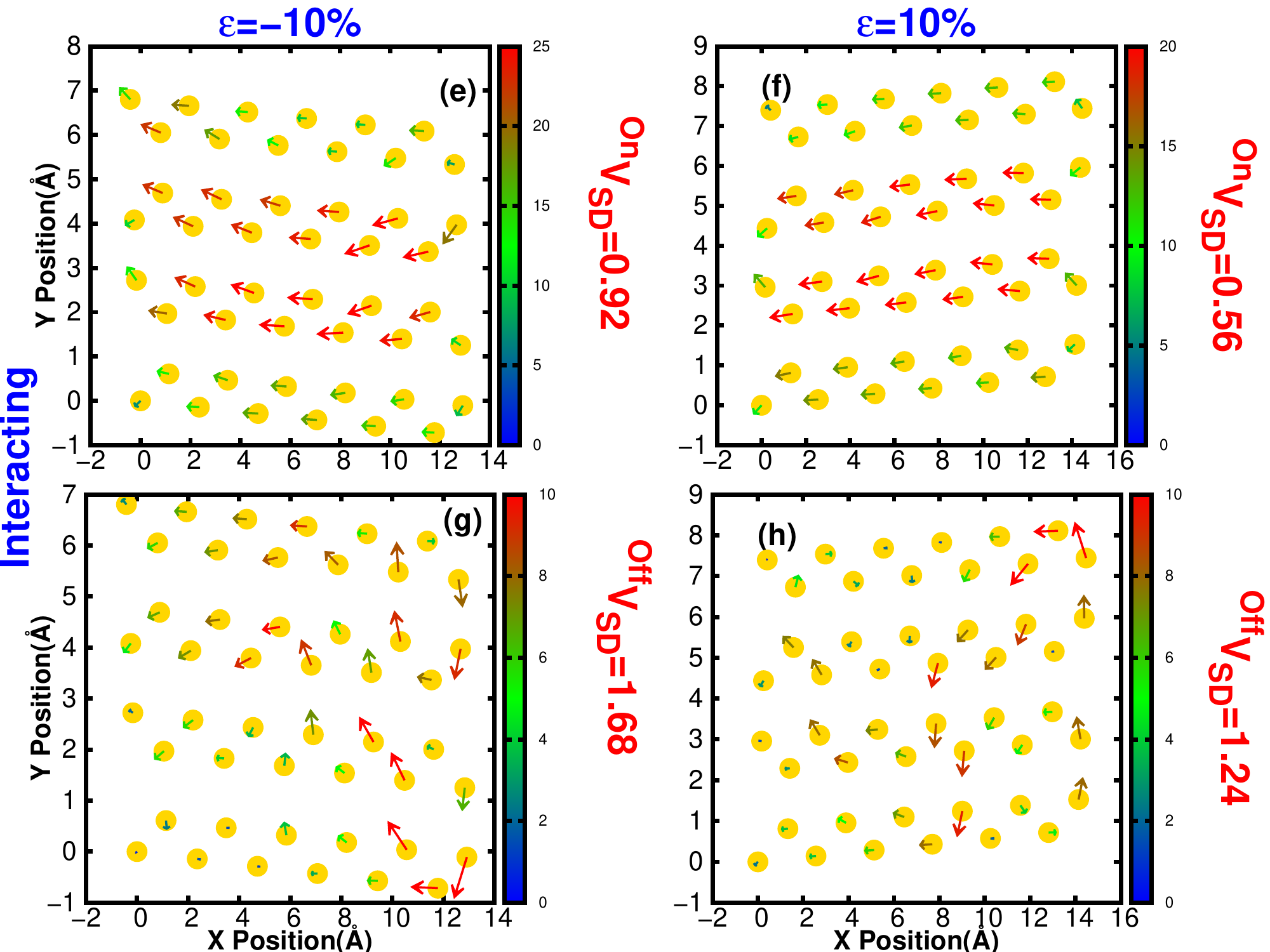}
\caption{Spatial profile of bond charge current through ZGNR(4,6) for the compressive ($\varepsilon=-10 \%$) (all panels in the left column) and tensile ($\varepsilon=+10 \%$) (all panels in the right column) strain along the $\theta=\pi/4$ in two cases; top panels (a-d) for non-interacting case, bottom panels (e-h) for interacting case. Here external bias is considered in the on-current voltage $^{on}V_{SD}$ (panels a,b,e,f) and also off-current voltage $^{off}V_{SD}$ (panels c,d,g,h). The gate voltage is $V_g=0.5$. The length of arrows is proportional to the resultant current on each site. The color-box displays bond current which is measured in the unit of $\mu A$.}
\label{bond_current}
\end{figure}
Fig. \ref{bond_current} shows spatial profile of local charge currents on each atomic site under application of compressive and tensile strain along $\pi/4$ direction. The top (a-d) and bottom (e-h) panels compare the effect of e-e interaction on spatial bond currents when strain is present. The source-drain bias is chosen in two situations: on-current state $^{on} V_{SD}$ and also off-current state $^{off} V_{SD}$.


Let us focus on {\it non-interacting} case (Fig.\ref{bond_current} a-d) in which the general feature points out that the local current has small values at the edges and large values in the middle of ZGNR. In other words, accumulation of charge at the edges causes to have small local currents at the edges. This is in agreement with the charge continuity equation (Eq.\ref{continuity}). For the compressive strain (Fig.\ref{bond_current} a), local current at the upper edge is more enhanced than the lower one, while as it is demonstrated in Figs. \ref{fig:compressivecharge} and \ref{fig:diffcharge}a, charge is much more accumulated on the lower edge than the upper one. All feature becomes inverse for the tensile strain (Fig.\ref{bond_current} b). Arrows which are indicator of the local currents are aligned mostly along the transport axis (zigzag chains). However, as voltage reaches to the off-current state, this alignment will be disturbed such that in the source-side, current tends to flow through those bonds that have large hopping parameters. As a result, the forward current will be decreased especially in the middle of ZGNR.     

In the {\it interacting} case, spatial profile of local current (Fig.\ref{bond_current} e,f) represents that resultant current at the on-current state has the same behavior as the {\it non-interacting} case. For both strains, current mostly passes through the middle of ZGNR than the edges. However, it is interesting that at the off-current state, transverse current emerges in the source-side as depicted in (Fig.\ref{bond_current} g,h). In fact, there is top-to-bottom current for the tensile strain and vice versa for the compressive strain. The reason comes back to charge difference accumulated on the edges which would be at its maximum values at the off-current voltage. Because of charge gradient between the upper and lower edges, transverse current flows to decrease such gradient of charge density. The described phenomena is the reason for having a decrease in off-current when e-e interaction turns on.

\section{Conclusion}
Strain can operate not only as an electro-mechanical switch of charge current, but also, by application of strain in different directions, one can control the direction of flowing local current among the nanoribbon. In this paper, charging effect of ZGNR is investigated in the presence of strain by using self-consistent non-equilibrium green's function formalism. The I-V characteristic curves show that negative differential resistance (NDR) and consequently its on-off ratio of the current strongly depend on the type of strain (tensile or compressive strain) and also the strength and direction of strain. The on-off ratio of current increases/decreases for the tensile/compressive strain. As an application proposal, instead of switching between on/off current by using the source-drain external bias, one can do switching by changing the strain type (for example from the tensile strain along $\theta=0$ direction to the compressive strain). For the oblique strain, parity conservation breaks out and as a consequence, some transport gaps are closed giving rise to affect $ I_{on}/I_{off} $ ratio which is drastically decreased in compared to the un-strained case. The electrostatic potentail induced by the tensile strain asymmetrically drops at the source side and would be flat in the bulk of ZGNR, while the compressive strain leads to symmetric potential drop at the contacts.     

Spatial profiles of the electrostatic potential, transferred charge and also bond local currents demonstrate that corresponding to the charge continuity equation, charge is accumulated on the nanoribbon edges while simultaneously, current flows mostly through the middle of ZGNR. Moreover, for the oblique strains, accumulated edge charge would be imbalanced and so, local current would be asymmetry across the nanoribbon. As a result, one can measure a transverse voltage appeared between the upper and lower edges of ZGNR. This transverse voltage increases by the strain strength.

 As a conclusion, there is a relation between imbalanced edge charge and longitudinal current through ZGNR. A minimum imbalanced charge at the edges leads to the on-current state while maximum imbalanced charge results in the off-current state. 
\appendix

\section{Quantum Transition Rules in Transport}
\label{appendix:a}
\begin{figure}[H]
\centering
\includegraphics[width=1\linewidth]{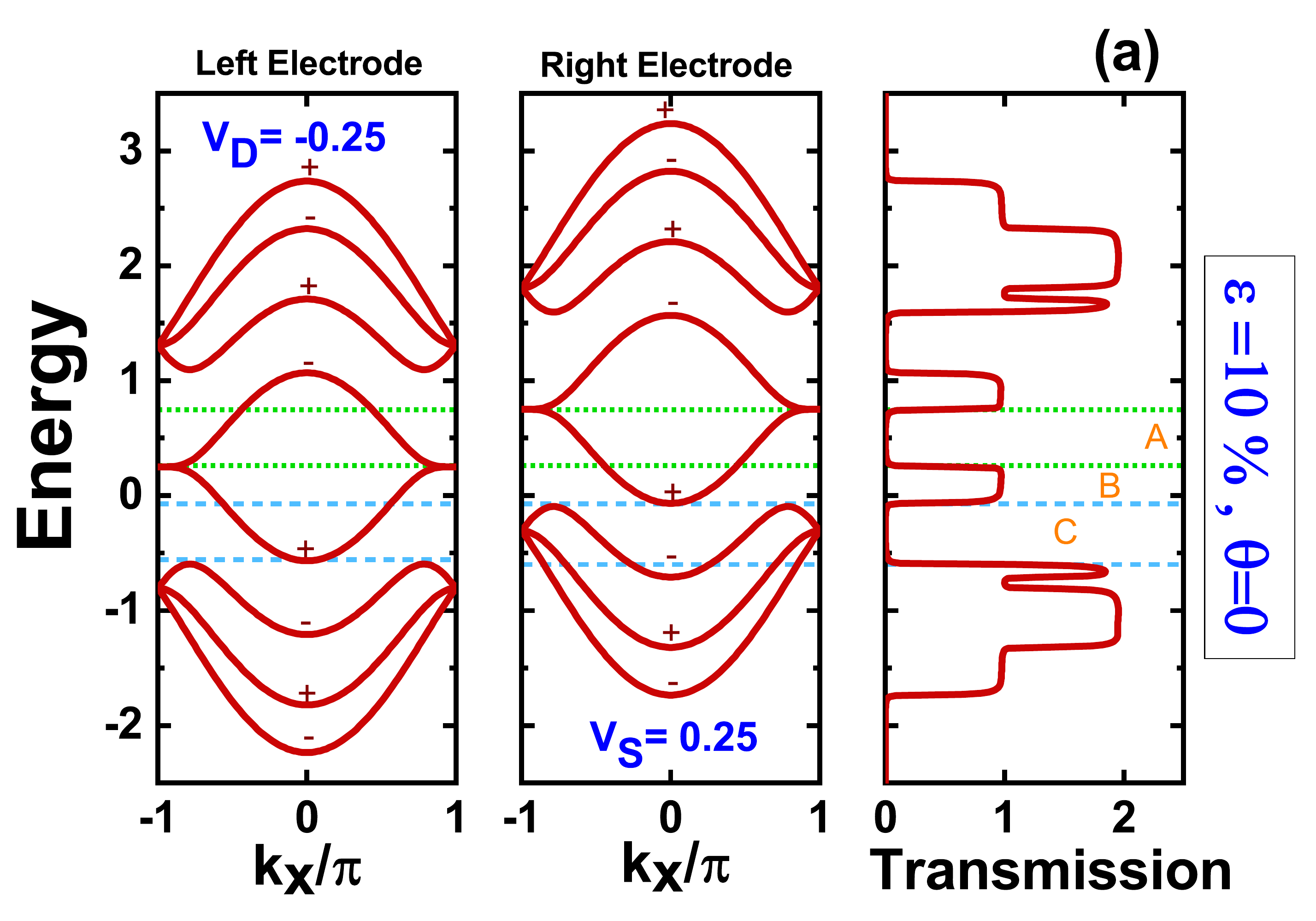} 
\includegraphics[width=1\linewidth]{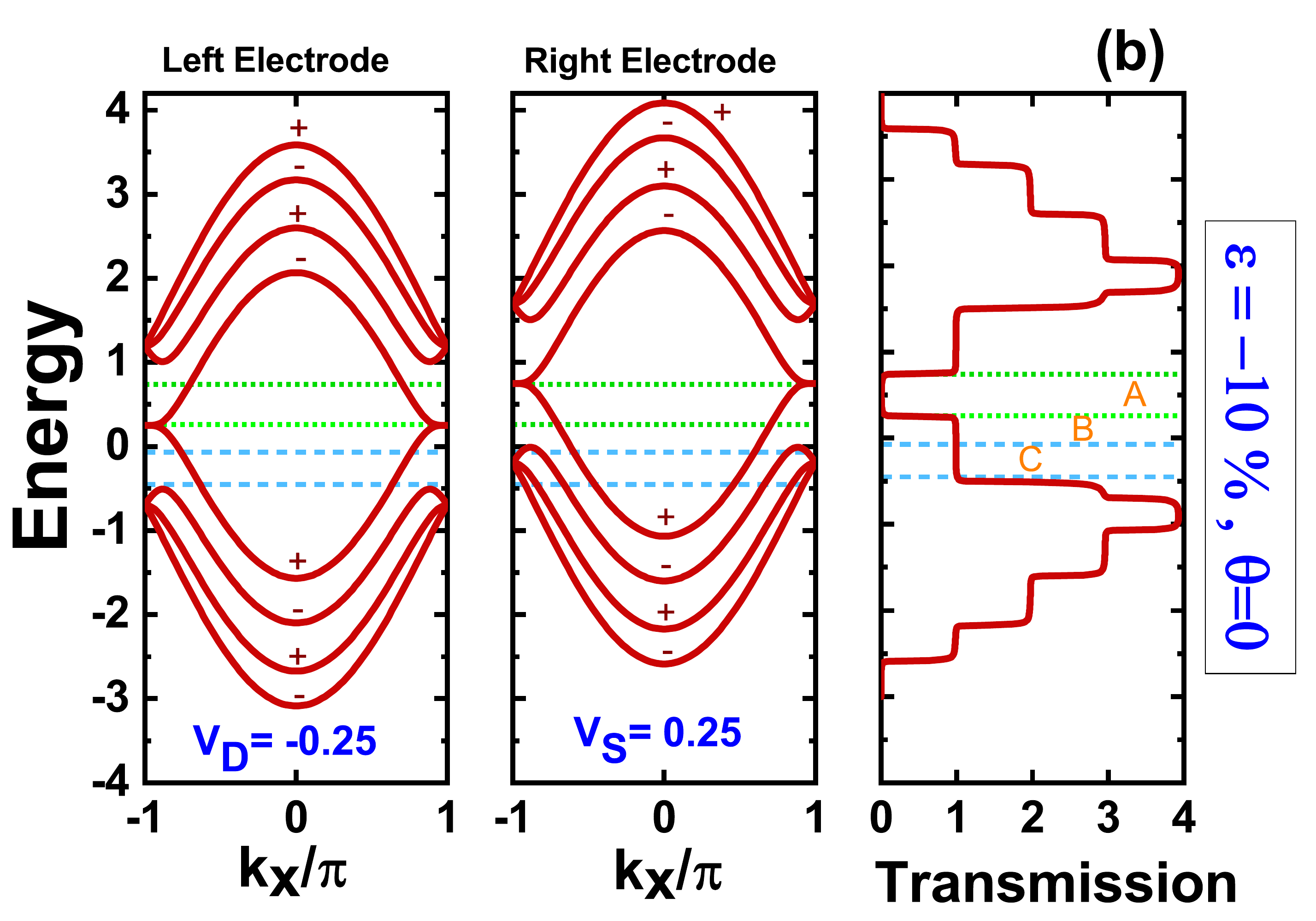}
\caption{Band structure of the left and right electrodes and also transmission through ZGNR(4,6) under applied gate $ [V_g=0.5] $ and source-drain bias $ [V_{SD}=0.5] $ (The first red dashed line from the left side of Fig.\ref{fig:current0-90}).The strain is directed to the $ \theta=0 $ and its strength would be as (a)$ \varepsilon=+10\% $ for the tensile strain  and (b) $ \varepsilon=-10\% $ for the compressive strain. The parity of each band is determined by +/- signs on band structures.}
\label{appfig1}
\end{figure}

In this appendix, we review the transition rules governing strained ZGNR with even number of zigzag chains in width. Since the mirror symmetry of ZGNR is conserved even if applied strain is directed along $\theta = 0$ or $\pi/2$, the bands are still labeled by their parity\cite{Grosso1,Grosso2,cheraghchi2010gate}. The band structure of the left and right electrodes and also transmission through ZGNR are shown in Fig. \ref{appfig1}, under applied a) tensile and b) compressive strain along the $\theta=0$ direction. For the tensile strain ($\varepsilon=10 \%$), there are two energy regions A and C which correspond to transport gap. The transport gap of A is originated from the transition between the states with opposite parity in the left and right electrodes. Based on the parity selection rule, this transition is forbidden. On the other hand, transport gap depicted by C is related to the second transition rule in which transition between disconnected bands (the central band to the lower bands group) is forbidden if the ribbon length would be enough long to prepare smooth variation of external bias. In a smooth variation of bias, electron momentum varies infinitesimally and as consequence electron is not scattered in the disconnected bands\cite{Grosso1,Grosso2,cheraghchi2010gate}. The energy region B corresponds to one conducting channel of the same parities in electrodes.

In the other case, let us focus on these energy regions of the compressive strain. The transport gap only happens in the energy region A coming from the parity conservation. However, there is one conducting channel in the energy region C which is an other evidence for prohibition of state transition between disconnected bands. A comparison between the tensile and compressive strain demonstrates that transmission in the compressive strain goes up to 4 in some energy ranges while in the tensile strain it reaches to 2 at most. Indeed for the compressive strain, the spectrum is much wider in compared to the spectrum of the tensile strain.

As a conclusion, the transition rules of ZGNR with even number of chains in width are preserved at the longitudinal and transverse strain.            

\end{document}